\documentclass[10pt,journal]{IEEEtran}
\usepackage{amsthm,amsmath,amssymb,array}
\usepackage[caption=false,font=footnotesize]{subfig}
\usepackage{arydshln}
\usepackage{mathrsfs}

\usepackage{hyperref}
\hypersetup{hypertex=true,
            colorlinks=true,
            linkcolor=red,
            anchorcolor=blue,
            citecolor=green}

\ifCLASSOPTIONcompsoc
  \usepackage[nocompress]{cite}
\else
  \usepackage{cite}
\fi
\usepackage{multirow}
\ifCLASSINFOpdf
  \usepackage[pdftex]{graphicx}
\else
  \usepackage[dvips]{graphicx}
\fi
\usepackage{algorithmic}

\begin{document}

\title{Joint Generative Learning and Super-Resolution For Real-World Camera-Screen Degradation}

\author{Guanghao~Yin, Shouqian~Sun, Chao~Li, Xin~Min
\IEEEcompsocitemizethanks{\IEEEcompsocthanksitem Guanghao Yin, Shouqian Sun, Xin Min, Chao Li are  with the Key Laboratory of Design Intelligence and Digital Creativity of Zhejiang Province, Zhejiang University, Hangzhou 310027.

E-mail: \{ygh\_zju, ssq, superli, minx\}@zju.edu.cn}
}

\markboth{Journal of \LaTeX\ Class Files,~Vol.~14, No.~8, August~2015}%
{Shell \MakeLowercase{\textit{et al.}}: Bare Demo of IEEEtran.cls for IEEE Transactions on Multimedia}

\IEEEtitleabstractindextext{%
\begin{abstract}
In real-world single image super-resolution (SISR) task, the low-resolution image suffers more complicated degradations, not only downsampled by unknown kernels. However, existing SISR methods are generally studied with the synthetic low-resolution generation such as bicubic interpolation (BI), which greatly limits their performance. Recently, some researchers investigate real-world SISR from the perspective of the camera and smartphone. However, except the acquisition equipment, the display device also involves more complicated degradations. In this paper, we focus on the camera-screen degradation and build a real-world dataset (Cam-ScreenSR), where HR images are original ground truths from the previous DIV2K dataset and corresponding LR images are camera-captured versions of HRs displayed on the screen. We conduct extensive experiments to demonstrate that involving more real degradations is positive to improve the generalization of SISR models. Moreover, we propose a joint two-stage model. Firstly, the downsampling degradation GAN(DD-GAN) is trained to model the degradation and produces more various of LR images, which is validated to be efficient for data augmentation. Then the dual residual channel attention network (DuRCAN) learns to recover the SR image. The weighted combination of L1 loss and proposed Laplacian loss are applied to sharpen the high-frequency edges.  Extensive experimental results in both typical synthetic and complicated real-world degradations validate the proposed method outperforms than existing SOTA models with less parameters, faster speed and better visual results. Moreover, in real captured photographs, our model also delivers best visual quality with sharper edge, less artifacts, especially appropriate color enhancement, which has not been accomplished by previous methods.
\end{abstract}

}

\maketitle
\IEEEdisplaynontitleabstractindextext
\IEEEpeerreviewmaketitle

\section{Introduction}
\label{Introduction}
\IEEEPARstart{T}{he} single super-resolution (SISR) is an elementary low-level vision task, which aims at the reconstruction of the high-resolution (HR) image from its low-resolution (LR) observation~\cite{glasner2009super}. The SISR has high practical values to enhance the quality of image to promote human visual experience, which has been applied in medical imaging~\cite{yang2012coupled}, satellite image enhancement~\cite{5702359} and facilitating other high-level tasks~\cite{gunturk2003eigenface-domain}.

The SISR is a seriously ill-posed inverse problem because of ill-conditioned registration, unknown degraded operators and multiple correspondence from a specific LR input to a crop of HR images~\cite{yang2010image,bertero1998introduction}. Generally, the researches of SISR focus on learning the pixel and texture prior informat ion from the paired HR and LR exemplar images~\cite{glasner2009super,kim2010single,freedman2011image,huang2015single}.

Existing SISR solutions can be divided into three types: interpolation-based methods, reconstruction-based methods and learning-based methods~\cite{yang2019deep}. Early interpolation-based solutions, such as bicubic interpolation~\cite{keys1981cubic} and Lanczos resampling~\cite{keysLanczos}, have the fast speed but produce yield poor results. Reconstruction-based solutions utilize complicated prior knowledge to restrict the reconstruction~\cite{dai2009softcuts,Single}. Learning-based solutions utilize machine learning models to mine the relationships from the LR-HR pairs. Since the classical SRCNN~\cite{dong2015image} has been proposed, deep convolutional neural network (CNN) based SISR methods are continually bringing prosperous improvement in terms of reconstruction accuracy~\cite{dong2016accelerating,shi2016real,kim2016accurate,lim2017enhanced,tong2017image,zhang2018image,dai2019second}.

However, the SISR research with complicated degradation still lacks effective exploration~\cite{yang2019deep}. Existing deep learning based methods are suffering limitations of generalization and robustness in real-world degradations~\cite{chen2019camera,zhang2019zoom,cai2019toward} because those models are well-designed for synthetic downsampling, such as bicubic interpolation (BI)~\cite{zhang2018learning}. For example, it can been seen in Fig~\ref{fig:sr_real} that the state-of-the-art models, ESRGAN~\cite{wang2018esrgan} and RCAN~\cite{zhang2018image}, are sensitive to the high-frequency Moir\'{e} pattern, although they have been trained with camera-screen degraded data. The popular SISR datasets with paired high-quality HR and artificial LR images lead to the over-fitting of DNN models on the certain degradation. There are two possible solutions that can be explored: (1) involving more LR images, which are more accordant with the complicated degradations in real-world conditions; (2) improving the representation ability of model to synthetically handle more complicated degradations. The recent trend of collecting real-world images~\cite{chen2019camera,zhang2019zoom,cai2019toward} and generating multiple simulated data~\cite{bulat2018learn,zhang2018learning,zhou2019kernel} is very positive, since it involves more degraded images and makes the resulting trained models performe better on real data.

\begin{figure}[t]
  \centering
  \includegraphics[scale=0.4]{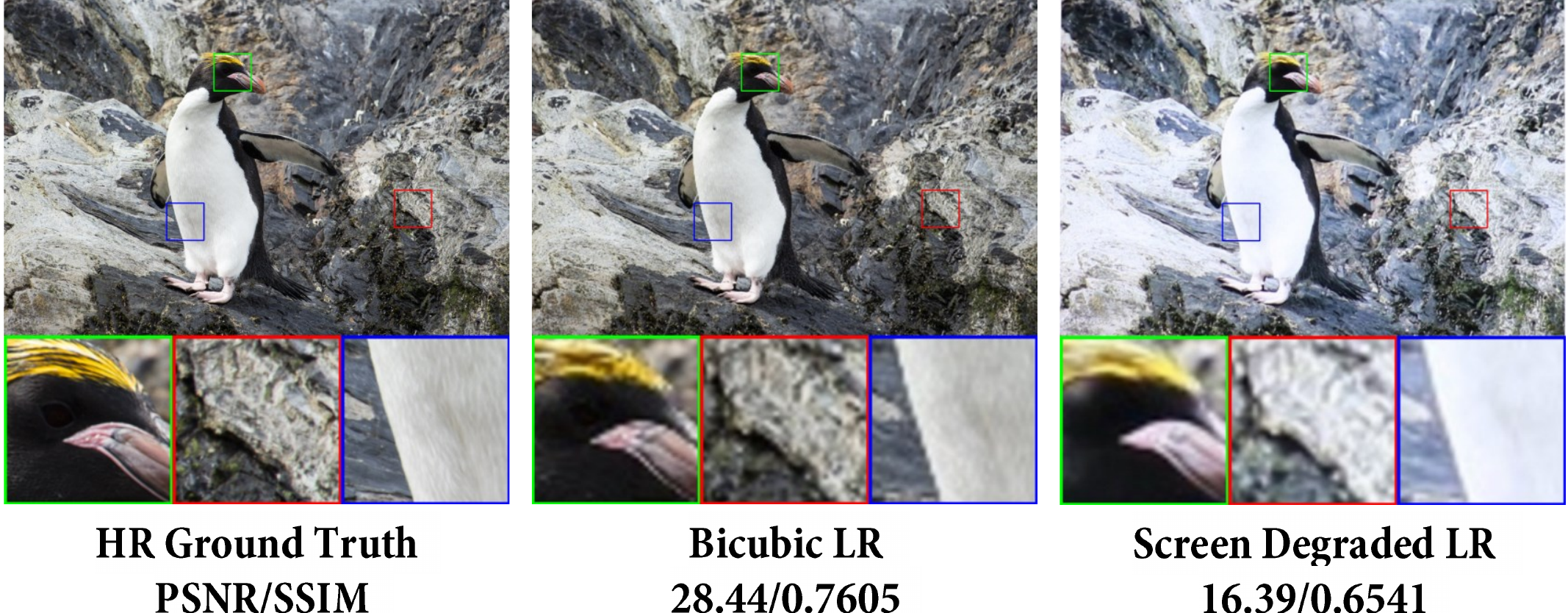}
  \caption{Visual comparisons between the LR image (X4) with bicubic downsampling and the camera-screen degradation (both are displayed after interpolation). The camera-screen example is much degraded with noise, blur, corruption and overexposure, which is quantitatively verified by PSNR/SSIM.
  }
  \label{example}
\end{figure}

 In this paper, we attempt to explore whether camera-screen degradation could effectively improve the performance and generalization of SISR models. Different from the film days, digital photos are directly shown by the display screen and people would like to use their image acquisition device to record contents on the screen for convenience. In this real-world scene, we found that the camera-screen degraded image was more complicated with noise, blur, corruption and overexposure under the joint influence of camera and screen, as shown in Fig.~\ref{example}. The estimation of camera degradation is non-uniform which cannot count on synthetic kernel estimation methods~\cite{cai2019toward,zhou2019kernel}. As the degradation is jointly influenced by camera and screen, the uniform solution becomes much more complicated. It should be characterized by obtaining real LR-HR pairs. Therefore, we establish a dataset named as Cam-ScreenSR, which contained the degradation from both the image acquisition and display device. The HR ground truths of the Cam-ScreenSR are from DIV2K dataset~\cite{timofte2017ntire}. And the LR images are the corresponding photographs captured from a monitor by the camera. At first, we just established the training/testing sets with the same camera and monitor. However, the same image appears significantly different when displayed and captured using different camera-display combinations~\cite{wengrowski2019light}, as shown in Fig.~\ref{fig:diff_screen}. Therefore, two more testing sets were collected with different equipment to validate that our data acquisition solution was not strict only to the specific screen and specific camera. It should be emphasized that our exploration focuses on the SISR task. Recovering the photographs from the camera-screen degradation is a more sophisticated task than what researchers usually call "pure super resolution". The existing SISR methods only restore an image that has been prefiltered by some kernel and then downsampled. It really limits the applications of SISR. The camera-screen degraded SISR task includes things like denoising, sharpening the edge, fixing color distortion, and so on, which each have a long history of study in image processing. However, if we attempt to improve the practicability of SISR solution in real-world scene, it's inevitable to involve complicated degradations. It should be encouraged, not strictly seperating research areas of image restoration.

To handle the camera-screen degradation, we also propose a joint generative learning and super-resolution model, as illustrated in Fig.~\ref{network}. The proposed model contains two networks: (1) The downsampling degradation GAN (DD-GAN) is used to learn the camera-screen degradation and generate more degraded LR images for data augmentation (the results shown in Fig.~\ref{fig:generateLR}). The DD-GAN focuses on overcoming the time-consuming and inefficient problem of large scale HR-LR manual acquisition. (2) The dual residual channel attention network (DuRCAN) recovers the mixed real captured and generated LR images. As existing pure SISR model can't handle complicated degradations, we involve the dual residual learning inspired by~\cite{liu2019dual}. The channel attention mechanism is applied to exploit the inter-channel relationship and adaptively reweights channel-wise features. We figure out that the dual residual blocks focus on recovering clearer textures and the channel attention blocks conduct the color calibration. Besides, similar to the Laplace operation commonly used in image processing~\cite{rosenfeld1976digital}, our solution additionally involves a Laplacian loss to sharpen the edge and smooth the noise.

Systemic ablation experiments have been conducted in the Cam-ScreenSR. And we compared our joint model with other SISR state-of-the-arts, which were all trained with the Cam-ScreenSR for fair comparisons. The comparisons were also conducted for the pure SISR task. The Cam-ScreenSR-trained models were finetuned with typical BI DIV2K training set and evaluated on popular SISR testing sets (Set5~\cite{bevilacqua2012low}, Set14~\cite{zeyde2010single}, BSD100~\cite{arbelaez2010contour}, Urban100~\cite{huang2015single}). Moreover, the restoration of real-world photographs proved that our model could appropriatly conduct color enhancement because of the camera-screen degraded data and channel attention mechanism, which has not been accomplished by previous SISR tasks. The excellent improvements in those experiments validate the great robustness and generalization of our solution. Compared with existing SOTA models, the proposed method can produce better visual results with less parameters and faster speed. The results also prove that involving more complicated degradation is helpful to boost development for SISR task.

In summary, the contributions of our paper are: \begin{itemize}
  \item First involving the camera-screen degradation and proposing a data acquisition strategy to establish the Cam-ScreenSR dataset, which is proved to be helpful for typical and real-world SISR tasks.
  \item Proposing a downsampling degradation GAN(DD-GAN), which learns the degradation from real-captured data and generates more LR images to replace the time-consuming manual acquisition.
  \item Proposing the dual residual channel attention network (DuRCAN), which is a controllable model to jointly restore the high-resolution details and enhance the color from the degraded images.
  \item Adding a Laplacian loss to sharpen the edge and smooth the noise.
\end{itemize}

\section{Related Work}
\label{Related Work}
\subsection{Deep Learning Based Single Image Super-Resolution.}
As a long-standing problem, early solutions for SISR task utilized the prior statistics~\cite{aly2005image,xiong2010robust,he2013beta} or exemplar patches~\cite{freeman2002example,glasner2009super}. Due to the superior performance of the pioneer SRCNN model~\cite{dong2015image}, deep learning-based methods have became the hotspots to tackle the ill-posed SISR problem. Then, researchers focused on designing deeper network structure with larger receptive field, such as VDSR~\cite{kim2016accurate}, DRCN~\cite{kim2016deeply}. To utilize hierarchical features from different layers, many recent models also apply residual connections and dense blocks to mine the different frequency information from weight layers, such as SRDenseNet~\cite{tong2017image}, EDSR~\cite{lim2017enhanced}, RCAN~\cite{zhang2018image}. After various novel architectures and training strategies  have been proposed, the SISR performance gets continuously improved, such as Peak Signal-to-Noise Ratio (PSNR) and Structural Similarity (SSIM) values.

Some researchers noticed that the PSNR-oriented solutions trended to output over-smoothed results without sufficient details~\cite{ledig2017photo,wang2018esrgan}. Therefore, several explorations have been conducted to pursue visually pleasing results. The milestones, such as SRGAN~\cite{ledig2017photo} and ESRGAN~\cite{wang2018esrgan}, combined the adversarial loss and GAN framework to optimize the model in a feature space instead of pixel space~\cite{johnson2016perceptual}. Hence, those perceptual-driven SISR models can produce more photo-realistic results, which visually contain more irregular noises to rich high frequency details.

However, all those aforementioned works are trained to restore the limited artificial degradation. As mentioned before, the LR image in real-world SISR suffers more complicated degradations and all these existing models trained on synthetic datasets has poor ability to handle them~\cite{chen2019camera,zhang2019zoom,cai2019toward}.

\subsection{Real-world Datasets for Single Image Super-resolution}
The synthetic datasets have been widely used for training and evaluating the SISR solutions, including Set5~\cite{bevilacqua2012low}, Set14~\cite{zeyde2010single}, BSD100~\cite{arbelaez2010contour}, Urban100~\cite{huang2015single} and DIV2K~\cite{timofte2017ntire}. However, the SISR models trained with simulated data deliver poor results when applied to real LR images~\cite{efrat2013accurate}. It really limits the practicability of SISR models for real applications.
To overcome the limitations of uniform downsampling, some recent works address on capturing paired LR and HR photographs in real-world scenes. To the best of authors' knowledge, only three real-world SISR datasets have been established. Chen \textit{et al.}~\cite{chen2019camera} employed camera/smartphone from the perspective of camera lenses and conducted data rectification to get aligned LR-HR paired CameraSR dataset. Zhang \textit{et al.}~\cite{zhang2019zoom} addressed on the optical zoom functionality of the camera to establish the SR-RAW dataset. Cai \textit{et al.}~\cite{cai2019toward} established a larger benchmark dataset (RealSR), where the LR-HR pairs on the same scene were captured by adjusting the focal length of a digital camera. Different from existing camera-based strategies, our Cam-ScreenSR is the first attempt from the perspective of the acquisition and display device. We also attempt to prove that involving more complicated camera-screen degradations is indeed valuable for SISR task.

\begin{figure}[t]
  \centering
  \includegraphics[width = 8.5cm]{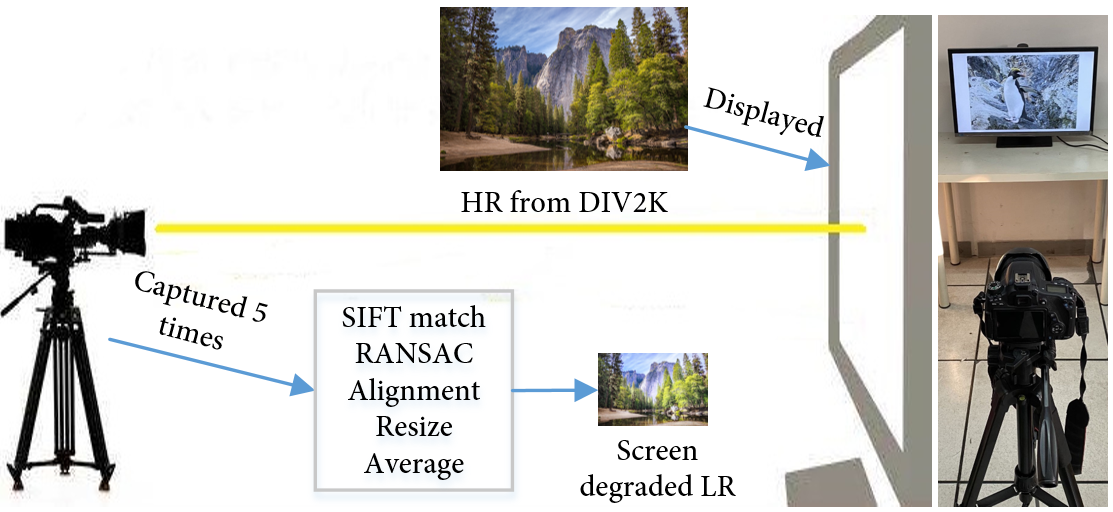}
  \caption{The calibrated acquisition equipment for the camera-screen degraded LR images. Without changing other conditions, the LR training set and one testing set are collected with Samsung S27R350 + Canon 760D camera. And the other two LR testing sets are collected with Dell IN2020M + iPhone 11, Lenovo X1 laptop + Huawei P30. The latter two testing versions are used to validate the generalization of our model, seen in Sention~\ref{exp}.
}
  \label{fig:position}
\end{figure}

\subsection{Residual Learning for Single Image Super-resolution}
The network depth is of crucial importance to the representation ability~\cite{simonyan2014very,szegedy2015going}. However, the stacked deep network suffers the notorious problem of vanishing/exploding gradients. After He \textit{et al.}~\cite{he2016deep} creatively proposed the concept of residual learning, the residual block became widely used in computer vision~\cite{krizhevsky2017imagenet,huang2017densely,he2017mask}. The residual connection provides a shortcut path to transfer the gradient of the error during back-propagation, which can effectively ease the training of deep networks (seen more details in~\cite{he2016deep}).

In SISR task, various researchers utilized the residual block as the basic unit to construct deep models for easier and more stable training ~\cite{tong2017image,lim2017enhanced,zhang2018image}. Kim \textit{et al.}~\cite{kim2016deeply} involved the residual skip-connection from input to the reconstruction layer, which could effectively supervise their recursive SISR model. Tong \textit{et al.}~\cite{tong2017image} utilized the dense residual connections to resue features from different layers and channels. Haris \textit{et al.}~\cite{haris2018deep} conducted iterative up-downsampling and used residual connections to project features. Recently, Liu \textit{et al.}~\cite{liu2019dual} proposed the concept of dual residual learning for noise removal, motion blur removal, haze removal, raindrop removal and rain-streak removal, where the dual residual connections provide more path to deliver features between the paired large- and small-size convolution kernels. The different combinations of dual kernels also provide various receptive fields for different resolution. Hence, we refer to the dual residual convolution operation to structure the basic block for camera-screen SISR task.

\subsection{Attention Mechanism}
In human proprioceptive systems, attention generally provides a guidance to focus on the most informative components of an input~\cite{hu2019squeeze-and-excitation}.
In neural networks, attention mechanism is effective to mine the long-range feature correlations in channel- and spatial-wise, which can guide models to reweight features and focus on more useful parts. Recently, the superiority of attention models has been proved in various tasks, ranging from image classification~\cite{hu2019squeeze-and-excitation,8100166,woo2018cbam} to language translation~\cite{vaswani2017attention}. As~\cite{zhang2018image} explains, in SISR task, the channel-wise features from different frequency are more informative for HR reconstruction. Therefore, we only involve the channel attention block to decrease the parameters of model. Moreover, we have explored that the channel attention can provide the ability of color calibration for our SISR solution with camera-screen degradation, seen in Section~\ref{exp}.

\section{Data Acquisition Strategy}
\label{Data Acquisition}
To capture realistic camera-screen degradation, we display the original images of DIV2K dataset on a Samsung S27R350 monitor. The resolution of screen keeps the maximum $1920 \times 1080$ with the 16:9 aspect ratio. To maintain the picture quality of the original source, the monitor is set to Standard mode (Brightness: 30, Contrast: 75, Sharpness: 64). The HR images are fullscreen displayed with Microsoft photo viewer. For image acquisition device, we utilize a DSLR camera (Canon 760D) to capture the camera-screen degraded LR images. The resolution of Canon 760D is $6000\times 4000$ and we capture LR observation at minimum 18mm focal length. Similar to the settings in~\cite{cai2019toward}, the camera is set to aperture priority mode and the ISO value is set to the lowest level to alleviate noise. The camera focuses on the center of monitor. The white balance and exposure are set to automatic mode.
\begin{table}[t]
  \begin{center}
  \centering
  \caption{Camera and screen specifications for Cam-ScreenSR dataset.}
  \label{Equipment}
  \centering
  \begin{tabular}{c|c|c|c}
  \hline
  \hline
  Cam-ScreenSR&Camera&Screen&Resolution \\
  \hline
  Training Set&Canon 760D&Samsung S27R350&$1920 \times 1080$\\
  testing set 1&Canon 760D&Samsung S27R350&$1920 \times 1080$\\
  testing set 2&Huawei P30&Lenovo X1&$3840 \times 2160$\\
  testing set 3&iPhone 11&Dell IN2020M&$1600 \times 900$\\
  \hline
  \hline
  \end{tabular}
\end{center}
\end{table}

Not only used for training, the Cam-ScreenSR training data also guides our DD-GAN to produce various degraded LRs. However, Camera properties (spectral sensitivity, radiometric function, spatial sensor pattern) and display properties (spatial emitter pattern, spectral emittance function) cause the same image to appear significantly different when displayed and captured using different camera-display hardware~\cite{wengrowski2019light}. Therefore, to validate that our solution is not strict only to the specific equipment, we added two more testing sets. The added display devices are Dell IN2020M ($1600 \times 900$) and Lenovo X1 laptop ($3840 \times 2160$), which are set to Standard mode similar to the Samsung S27R350 monitor. And we capture two versions of testing sets by the smartphone iPhone 11 and Huawei P30. Referring to~\cite{cai2019toward}, the configurations of the smartphone are similar to that for DSLR camera by using the ProCam software. To avoid less-effective repetition, we just present the training data acquisition strategy in the follow.
\begin{figure}[t]
  \centering
  \includegraphics[width = 8.5cm]{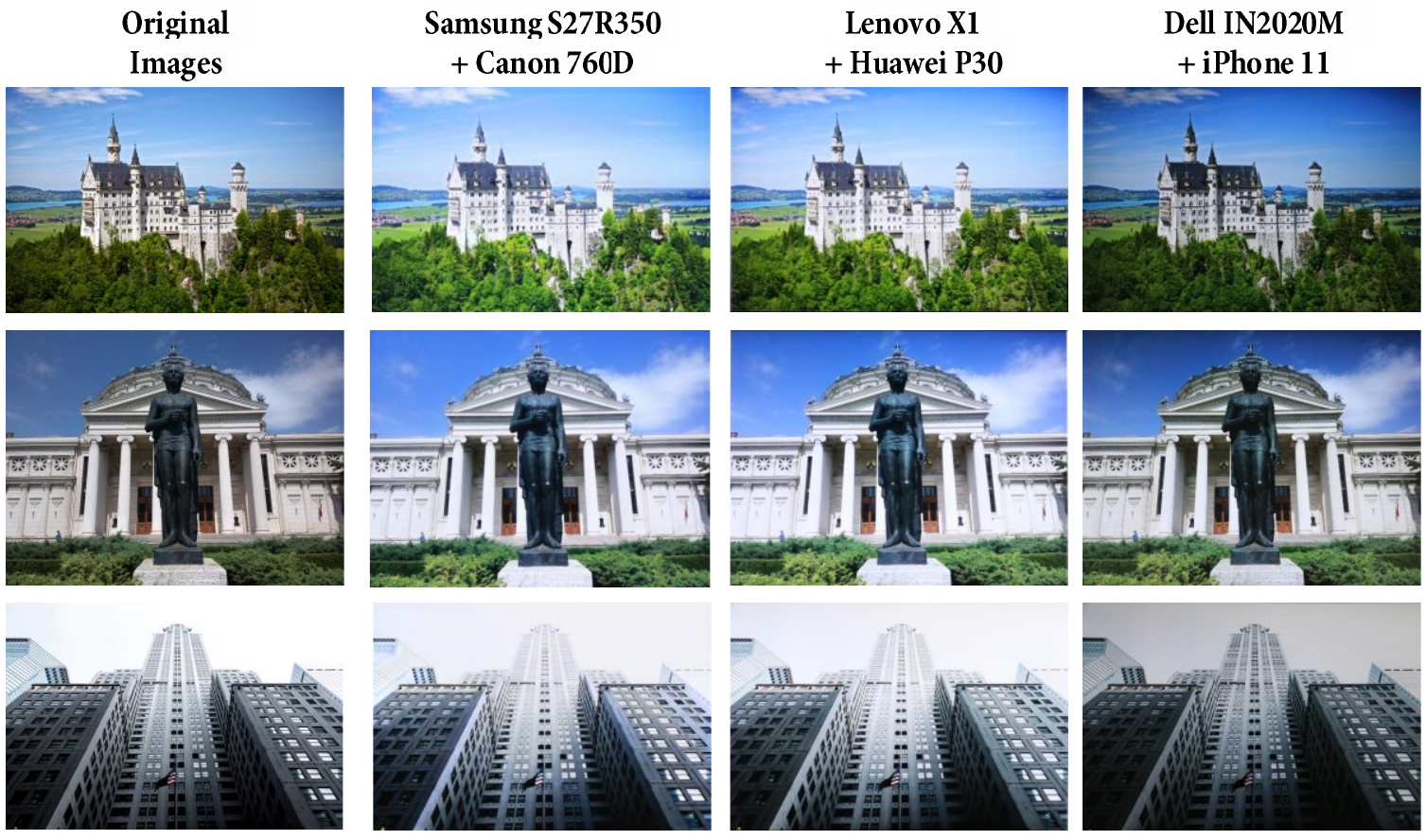}
  \caption{Cam-ScreenSR examples: Our dataset contains one camera-screen degraded training set and three testing sets. Each column corresponds to a different camera-display pair. The same image appears significantly different with different camera-display combinations. (Best viewed as zoomed-in PDF.)
}
  \label{fig:diff_screen}
\end{figure}

As shown in Fig.~\ref{fig:position}, the monitor is put in front of the clear background and the camera is mounted on a stable tripod at a distance of about 1 meter from the screen. To minimize the spatial misalignment and lens distortion, we utilize the mapping equipment to calibrate the camera lens parallel to the monitor and adjust camera to the same height of the screen center. For keeping stabilization, the camera connection software is used to control the shutter remotely. Different from zooming the camera lens in~\cite{cai2019toward,chen2019camera,zhang2019zoom}, our image collection strategy maintains the focal lengths which can avoid lens distortions at different focal lengths. Therefore, the fixed focal length and calibrated spatial position make our image pair alignment more easier. Each HR image is five continuous captured. Similar to CameraSR~\cite{chen2019camera}, we conduct SIFT key-points match~\cite{lowe2004distinctive} between the original HR images and the five captured LRs. Then, we utilize RANSAC~\cite{fischler1981random} to filter mismatched coordinates and estimate the homography. After getting the alignment parameters, we obtain five aligned images and average them to the one. Finally, the interpolation with scale factor is conduct to produce low resolution images. According to~\cite{chen2019camera}, the smoothing effects from the interpolation is not critical for LR images. It this paper, we typically choose the scale factor 4 to avoid the less-effective experiment repetition, similar to previous articles~\cite{ledig2017photo,wang2018esrgan}.

For data rectification, we conduct cropping, alignment, interpolation, but not luminance adjustment. The intensity variation is an important characteristic of camera-screen degradation. In practice, when facing LED shiner from different monitors, the aperture was auto-enlarged. Then, the captured photographs have different degrees of exposure and color distortion (as shown in Fig.~\ref{fig:diff_screen}). This situation should not be ignored because it's consistent with human pupil dilation. Moreover, the camera-screen degraded data with color distortion plays an important role to the PSNR/SSIM indexs and also guides the model for color correcting, which greatly expands the practicability of our solution (seen in Section.~\ref{exp}).

\begin{figure}[t]
\centering
\includegraphics[width=8.9cm]{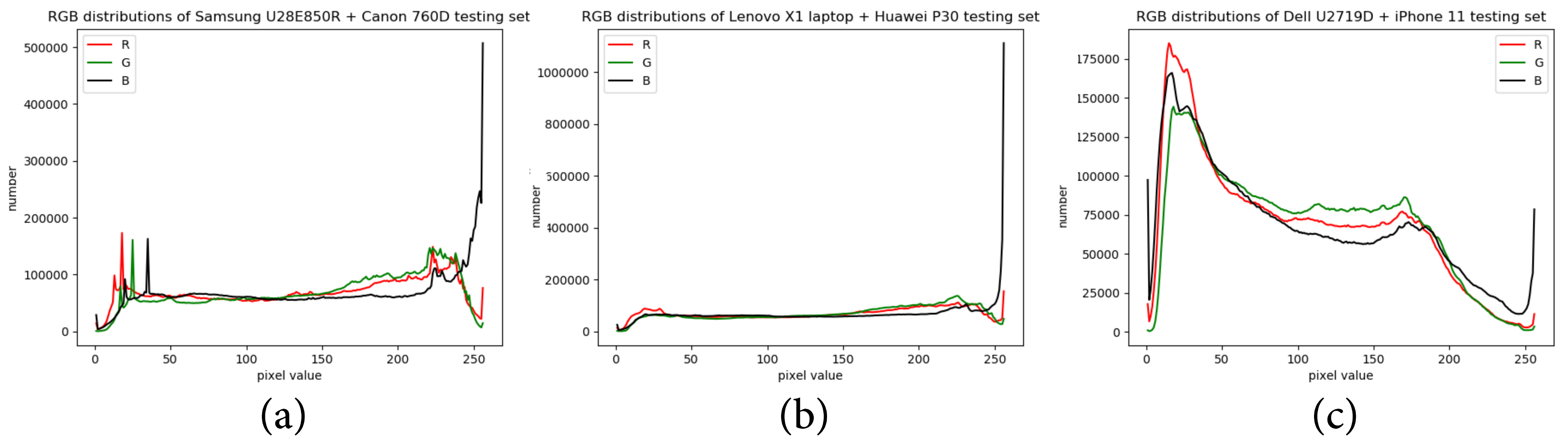}
\caption{The distributions of RGB channels of three Cam-ScreenSR testing sets. The figures in (a), (b), (c) separately correspond to the Samsung S27R350 + Canon 760D, Dell IN2020M + iPhone 11 and Lenovo X1 laptop + Huawei P30 testing datasets. }
\label{RGBdis}
\end{figure}

We define the data acquired by Samsung + Canon 760D, Lenovo X1 + Huawei P30 and  Dell + iPhone 11 combinations as the testing set 1, 2, 3. The RGB channel distributions of three testing sets are illustrated as Fig.~\ref{RGBdis}. Compared to testing set 3, the distributions of testing set 1 and 2 versions are relatively similar, because those two monitors have higher resolution and better color revivification degree. When LED shiners of the monitor are more luminous than the environment, the aperture of the camera will be auto-enlarged to receive more light. It explains the images of testing set 1 and 2 are lighter than HR images, and the one of testing set 3 is darker, as illustrated in Fig.~\ref{fig:diff_screen}.

\begin{figure*}[ht]
  \centering
  \includegraphics[width=18cm]{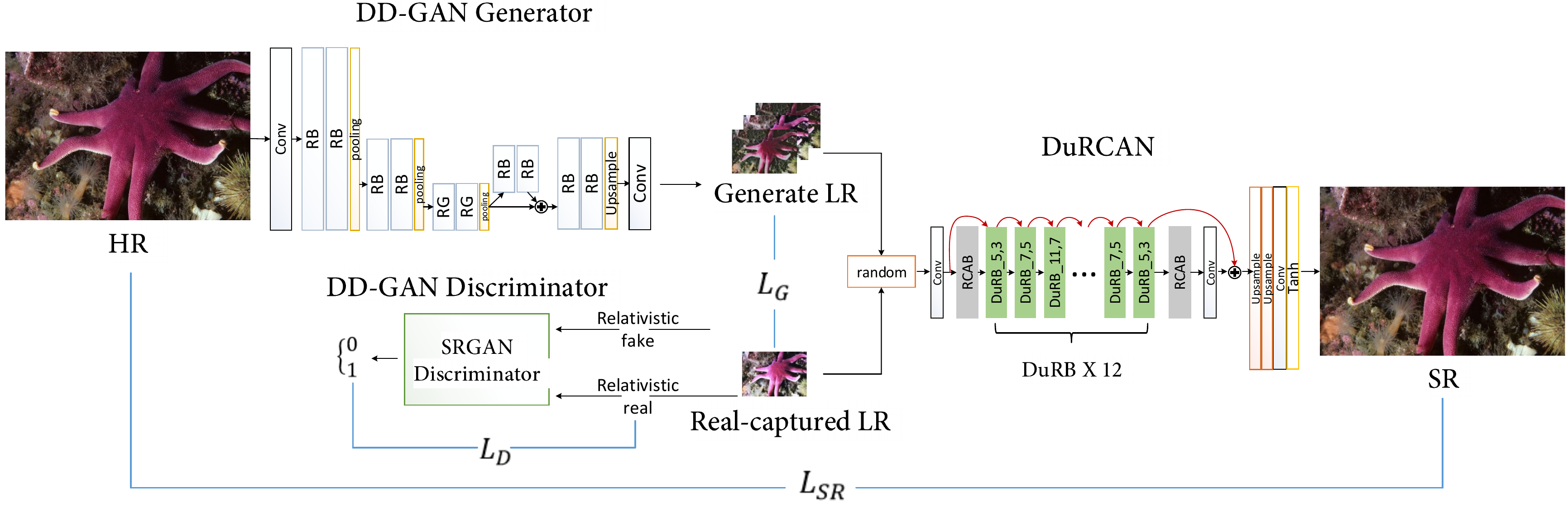}
  \caption{Overview of the proposed model. The number of upsample blocks can be adjusted for other scale factors. After joint training, the parameters of DD-GAN and DuRCAN can be finetuned for other situations.
}
  \label{network}
\end{figure*}

\section{Proposed Network}
\label{proposed network}
\subsection{Overview}
As described in Section~\ref{Data Acquisition}, we have established the camera-screen degraded super-resolution dataset (Cam-ScreenSR), which consists of paired LR-HR images $\{Y,X_{LR}\}$. To avoid the less-effective experiment repetition, our dataset focuses on SISR with scale factor 4. The size of $N$ HR ground truths ($Y=\{Y_1,...,Y_n\}$) is $h\times w$, and the paired LR ($X_{LR}=\{X_{LR1},...,X_{LRn}\}$) is $\frac{h}{4}\times\frac{w}{4}$. Previous SR formulations only consider the influence of camera~\cite{zhang2018learning,zhou2019kernel,cai2019toward}.The camera-screen degradation $D_{SR}(\cdot)$ is a comprehensive function with noise, blur, luminance corruption and downsampling from acquisition and display device. As expressed in~\cite{efrat2013accurate}, we can define the camera-screen degradation $D_{SR}(\cdot)$ as:
\begin{align}
  X_{LR} &= D_{SR}(Y)\\
  &= ((f(Y\ast k1)\downarrow^1 + n_1)\ast k2)\downarrow^2 + n_2,
\label{eq2}
\end{align}
where $f(\cdot)$, $k_1$, $\downarrow^1$, $n_1$ denote the image distortion, degraded kernel, downsampling operator, additive noise arose from the screen respectively and $k_2$, $\downarrow^2$, $n_2$ are those from the camera. Obviously, simplifying the noise and downsampling components like previous blue-kernel estimated SISR methods~\cite{zhang2019deep,gu2019blind} will underestimate the camera-screen degradation. Considering it is difficult to derive an numerical solution for Eq.~\ref{eq2}, we directly learn the SR restoration with real camera-screen degraded photographs.

The overall architecture of our model is illustrated in Fig.~\ref{network}. The data collection with various combinations of camera/monitor is a huge workload. Therefore, we utilize the generative learning to simulate camera-screen degradation from limited real-captured data and conduct data augmentation. The HRs ($Y=\{Y_1,...,Y_n\}$) from training set are fed into the DD-GAN. The generator $G_{\Theta_1}$ produces generated LR images ($X_{GLR}$) as:
\begin{align}
  X_{GLR} = G_{\Theta_1}(Y).
\end{align}
Referring to the Relativistic GAN~\cite{jolicoeurmartineau2019the}, the discriminator $D_{\Theta_2}$ predicts the probability that a real LR image $x_{LR}$ is relatively more realistic than the average of generated fake images $x_{GLR}$, which guides the generator to produce more realistic outputs. Following Goodfellow \textit{et al.}~\cite{goodfellow2014generative}, the DD-GAN is optimized to solve the adversarial min-max problem:
\begin{align}
  \min\limits_{\Theta_1}\max\limits_{\Theta_2}\ &\mathbb{E}_{Y\sim p_{train}(Y)}[\log( D_{\Theta_2}(X_{LR},G_{\Theta_1}(Y)))]+\notag\\
  &\mathbb{E}_{X_{LR}\sim p_{G}(Y)}[\log (1- D_{\Theta_2}(G_{\Theta_1}(Y),X_{LR}))].
\end{align}

Then, the real captured LR $X_{LR}$ and generated $X_{GLR}$ are randomly sent to the SR restoration network DuRCAN. We can obtain the SR images $\hat{Y}$ as:
\begin{align}
  \hat{Y} = S_{\Theta_3}(X),
  X = \{X_{LR},\gamma X_{GLR}\},
  \label{eq3}
\end{align}
where $\gamma$ is the mixing rate.

It should be emphasized the DD-GAN and DuRCAN are jointly trained where the failure outputs from the unbalanced generator can promote the robustness of DuRCAN. After training, the DuRCAN is used to restore LR images.

\subsection{High-to-Low Downsampling Degradation GAN}
The DD-GAN conducts a high-to-low generating, which simulates the process of camera-screen degradation to get more synthetic LR images for data augmentation.
Previous work ~\cite{bulat2018learn} in face super-resolution learns the artificial degradation pattern by concatenating noise vectors with unpaired HR images which easily causes mode collapse. Different from the aforementioned model, our DD-GAN directly utilizes real captured image pairs.

\subsubsection{Generator}
The generator relies on a encoder-decoder architecture for downsampling degradation. Given the input HR images $Y$, the generator of DD-GAN output the synthetic LR images $X_{GLR}$ to augment the limited real captured LR images. The downsampling and degradation process is modeled as two-stages: (1) the contracting subnet encodes the features of HR inputs. (2) the expansive subnet decodes the internal features from the contracting subnet to inversely generate camera-screen degraded LR images.

\begin{figure}[t]
  \centering
  \includegraphics[width = 6cm]{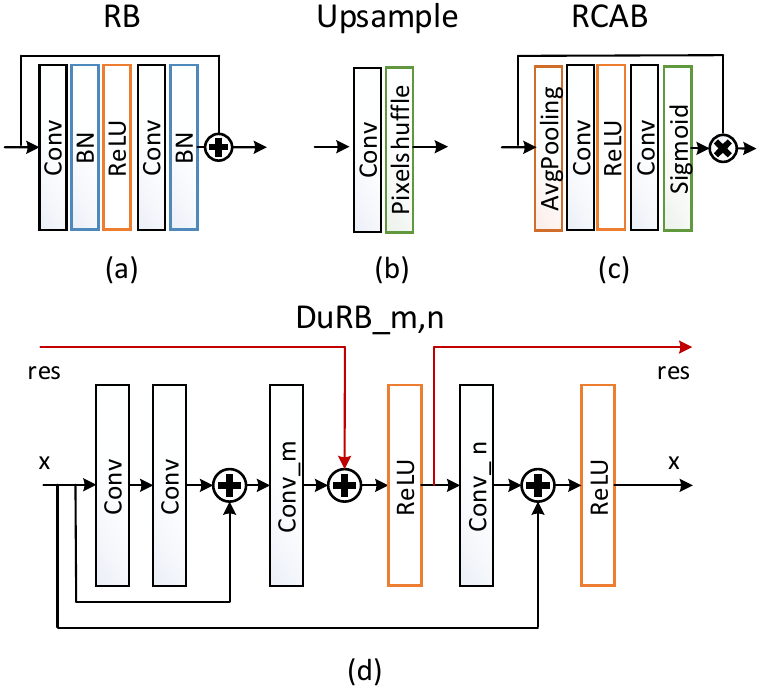}
  \caption{The unit architectures of (a) the residual block (RB); (b) upsample block (Upsample); (c) residual channel attention block (RCAB) and (d) dual residual block with the small- and large-kernel sizes $[T_m,T_n]$ (DuRB\_m,n).}
  \label{fig:DuRB}
\end{figure}

As illustrated in Fig.~\ref{network} and Fig.~\ref{fig:DuRB} (a), the typical Res-block~\cite{he2016deep} with MaxPooling operation is the stacked unit. Firstly, the single convolutional layer extracts shallow features from the input HR images. Then, those features are processed by the contracting subnet, which consists of 3 repeated groups. Each residual group contains two Res-blocks followed by a $2\times2$ MaxPooling operation for downsampling. Although maxpooling is not recommended in SR task for reducing image details~\cite{kim2016deeply}, it's suitable in the reverse high-to-low downsampling degradation~\cite{bulat2018learn}. Two stacked Res-blocks conduct "bottom" feature extraction. After that, the expansive subnet decodes concatenated features from previous layers and corresponding layers of contracting subnet with the PixelShuffle upsample-blocks~\cite{shi2016real}, the number of which is calculated by $N-log_2S$ where N is number of contracting layers and S is the scale factor. We attempt to output $\times4$ LR images. Hence, the contracting subnet consists of one upsample-block and one convolutional layer to get 3-channel output.

\subsubsection{Discriminator}
The discriminator of standard GAN estimates the probability that the input image is real, which guides the generator to  increase the probability that fake data is as real as ground truth. However, our DD-GAN tries to generate more various degraded images during the generative learning. If we apply standard GAN, the output will fall into the specific degradation pattern similar to the Cam-ScreenSR training data. Therefore, we enhance the discriminator with the relativistic label~\cite{jolicoeurmartineau2019the}, which can effectively decrease the realistic probability of real data during the training. The average evaluation of relativistic discriminator predicts the probability that a real image is relatively more realistic than all fake data in a batch, which formulated as
\begin{align}
  D_{\Theta_2}(X_{LR_i},X_{GLR})\rightarrow 1,\\
  D_{\Theta_2}(X_{GLR_i},X_{LR})\rightarrow 0.
\end{align}

\begin{figure}[t]
  \centering
  \includegraphics[width = 8.9cm]{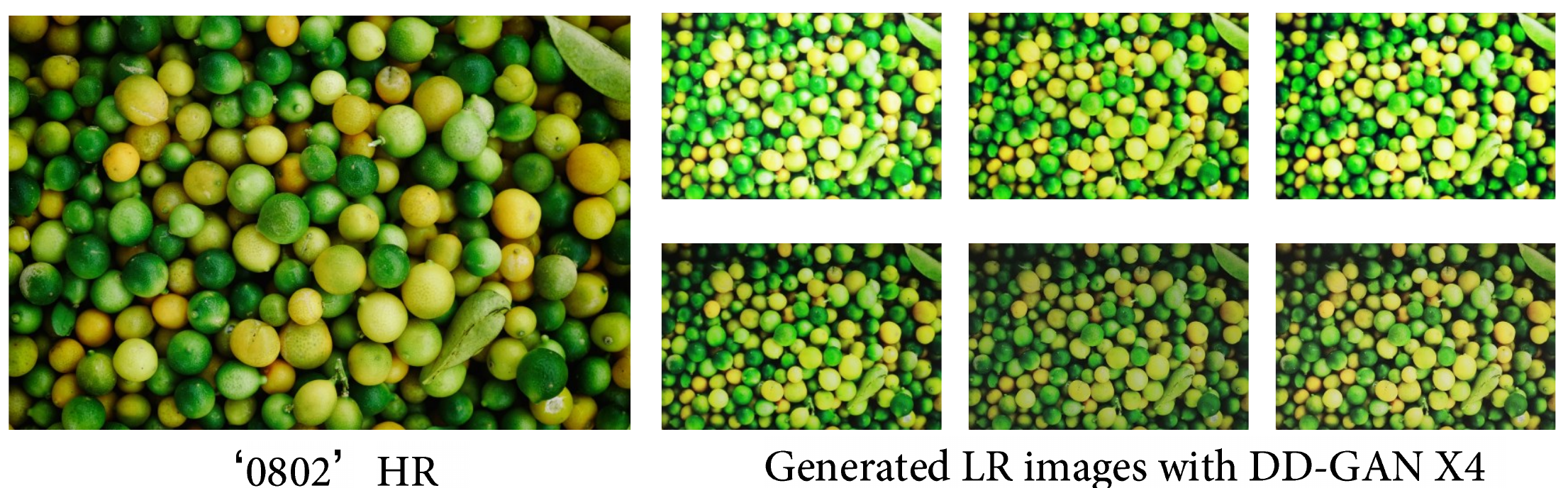}
  \caption{Examples of LR images produced by our DD-GAN trained with real-captured Cam-ScreenSR training dataset. It can be seen that the DD-GAN tends to generate more varieties of degradation which is benefit for improving the performance of SR models (described in Section~\ref{exp}).
}
  \label{fig:generateLR}
\end{figure}

Specifically for our task, we follow the architecture of SRGAN discriminator ~\cite{ledig2017photo} and enhance it with the average evaluation from relativistic discriminator~\cite{jolicoeurmartineau2019the}. The paired fake generated data $X_{GLR}$ and real-captured data $X_{LR}$ are fed into the SRGAN discriminator $C(\cdot)$ to predict the probability. Then the output of target type subtracts the  average of the opposite type in the mini-batch, followed with a Sigmoid function, which is formulated as:
\begin{align}
  D_{\Theta_2}(X_{LR},X_{GLR}) = \delta(C(X_{LR})-\mathbb{E}_{\mathbb{Q}}[C(X_{GLR})]),\\
  D_{\Theta_2}(X_{GLR},X_{LR}) = \delta(C(X_{GLR})-\mathbb{E}_{\mathbb{P}}[C(X_{LR})]),
\end{align}
where $\delta(\cdot)$ is the Sigmoid function and $C(\cdot)$ is the SRGAN discriminator and $\mathbb{E}$ represents the average operation of the data in the mini-batch and $\mathbb{P,Q}$ respectively denote the distribution of real and fake data. The examples of generated LR images produced by the DD-GAN are shown in Fig.~\ref{fig:generateLR}.

\subsection{Low-to-High Dual Residual Channel Attention Network}
Previous CNN-based SISR approaches~\cite{dong2015image,kim2016accurate,kim2016deeply,tong2017image,zhang2018residual,zhang2018image} have achieved impressive results on synthetic datasets. However, those models have the poor ability to handle complicated degradation patterns in real-world (seen in Fig.~\ref{fig:screen_compare} and Fig.~\ref{fig:sr_real}), which motivates us to propose a novel low-to-high model with great generalization and robustness.

As shown in Fig.~\ref{network}, the DuRCAN consists of five components: the shallow feature extraction, the deep feature extraction based on dual residual group (DuRG), two residual channel attention blocks at the beginning (RCAB\_bg) and the end (RCAB\_ed) of DuRG and the final upscale reconstruction to output the SR image. The number of channels is set to 64 for the internal layers.

Firstly, one convolutional layer $H_{SF}(\cdot)$ extracts shallow feature $F_{SF}$ from the LR input $X$ as:
\begin{align}
  F_{SF}=H_{SF}(X).
\end{align}

Then, the residual channel attention block (RCAB\_bg)  reweights the shallow feature and focuses on more useful parts in channel-wise dimensions. As Fig.~\ref{fig:DuRB} (c) shows, the average-pooling operation firstly aggregates the channel information to get the channel average-pooled feature $F_{avg}^{h/4\times w/4\times c}$. Then, the descriptor is fed into a multi-layer perceptron (MLP) with two convolutional layers and a ReLU function. In order to reduce parameter overhead, the convolutional layer $W_0$ conducts channel reduction, where the channels of processed features are decreased by the reduction ratio $\gamma$ to $F_{avg_0}^{h/4\times w/4\times c/\gamma}$. And the latter layer $W_1$ recovers the feature shape. The sigmoid activation function $\delta(\cdot)$ generates the normalized channel attention weight $W^{h/4\times w/4\times c}$ between 0 to 1. After that, the original shallow feature is reweight by the multiplication with the channel attention weight. Overall, the reweighted channel-wise feature $F_{CA1}$ is computed as:
\begin{align}
  F_{CA1} &= \delta(MLP(Pool(F_{SF})))*F_{SF}\notag\\
         &= \delta(W_1(W_0(F_{avg})))*F_{SF}.
\end{align}

Later, the dual residual group focuses on the SR reconstruction, which consists of 6 dual residual blocks (DuRB). As illustrated in Fig.~\ref{fig:DuRB} (d), the DuRB receives features and residual components $[x_i,Res_i]$ from previous layer and outputs the processed $[x_{i+1},Res_{i+1}]$ to the next DuRB. It consists of two parts: (1) The residual unit with two stacked convolutional layers $C(\cdot)$ processes feature $x_i$ from the previous $i$th DuRB, which is formulated as $x_{ci}=C(x_i)+x_i$. (2) Two paired convolutional layers $[C_m(\cdot),C_n(\cdot)]$ focus on the SR reconstruction with different kernel sizes $[T_m,T_n]$. The ReLU operators are followed after two layers. Without a pyramid structure~\cite{lai2017deep}, the alternate kernels provide the different receptive fields to conduct reconstruction. Moreover, the alternate convolutions with large and small kernel also conduct coarse- and fine-grained feature extraction from multi-degraded LR images. During conducting the convolution, the dual residual connections not only involve the residual messages $Res_i$ from previous DuRB, but also provide paths to deliver features from the first kernel to the latter one and deliver residual components to the next DuRB. It can greatly increases potential interactions between each block. The whole processing can be formulated as:
\begin{align}
  &Res_{i+1} = C_m(C(x_i)+x_i)+Res_i,\\
  &x_{i+1} = C_n(C_m(C(x_i)+x_i)+Res_i) + x,
\end{align}
where the ReLU function is omitted in the equations. We set the kernel size of 12 DuRBs $\{[T_1^l, T_1^s],...,[T_{12}^l, T_{12}^s]\}$ as \{[5,3], [5,3], [7,3], [7,5], [11,5], [11,7], [11,7], [11,5], [7,5], [7,3], [5,3], [5,3]\}. As~\cite{lim2017enhanced,wang2018esrgan} point out, when the distributions of the training and testing sets differ a lot, batch normalization (BN) tends to introduce unpleasant artifacts and limit the generalization ability. Therefore, we don't involve the BN operation for our network.

Before upsampling to the original size of HR image, the reconstructed features are reweighted by another residual channel attention block (RCAB\_ed) and processed by one convolutional layer $C_{ed}(\cdot)$. Finally, after adding the 12th residual component $Res_{12}$ from the last DuRB, the reweighted feature $F_{CA2}$ are up-sampled using PixelShuffle $PS(\cdot)$~\cite{shi2016real}. And the last convolution layer $C_{la}(\cdot)$ followed with Tanh function $Tanh(\cdot)$ outputs the 3-channel SR image. The output SR image is computed as:
\begin{align}
  \hat{Y}= Tanh(C_{la}(PS(C_{ed}(F_{CA2})+Res_{12}))).
\end{align}

Moreover, in ablation experiments, we find the DuRBs pay attention on restoring the details from complicated degradations. As the depth of DuRB increases, the SR image contains clearer textures and less artifacts. The channel attention blocks (RCAB\_bg and RCAB\_ed) focus more on color calibration. Removing the RCABs has acceptable influence on the SR definition, but greatly limits the ability of color calibration. Hence, for different scenes, the DuRCAN can be finetuned specifically. With the support of sufficient computing resources, the deeper DuRG or other novel model can be involved to improve the SR definition. And the targeted finetuning of the RCAB can control the ability of color enhancement.

\subsection{Loss Functions}
\label{Loss Functions}
In this section, we will describe the loss functions in details.

\subsubsection{DD-GAN Loss Function} To avoid overfitting in the specific degradation pattern, we apply the label smoothing~\cite{muller2019does} for discriminator. The distinguishing labels of real and fake data are not static as 1 and 0, but randomly sampled from the uniform distribution $U(0,\alpha)$ and $U(\beta,1)$, where $\alpha$ and $\beta$ are near 0 and 1. The BCE loss $L_{BCE}(\cdot)$ is set to evaluate the distance between distinguishing label and predicted probability from the discriminator. Therefore, the parameters of DD-GAN discriminator is optimized by discriminator loss $L_D$ as:
\begin{align}
    L_D =& \mathbb{E}_{\mathbb{P}}[L_{BCE}(a,D_{\Theta_2}(X_{LR},X_{GLR}))]+\notag\\
    &\mathbb{E}_{\mathbb{Q}}[L_{BCE}(b,D_{\Theta_3}(X_{GLR},X_{LR}))],
\label{eq:discriminator loss}
\end{align}
where $\mathbb{P,Q}$ respectively denote the distribution of real and fake data, and $a,b$ are the random values in the range of $U(0,\alpha)$ and $U(\beta,1)$.

For generator, the loss function $L_G$ is the combination of the content loss $L_{con}$ and adversarial loss $L_G^a$. The content loss $L_{con}$ consists of a perceptual loss $L_{vgg19\_54}$ and a pixel-wise loss $L_1(\cdot)$, which is consistent with the previous ESRGAN~\cite{wang2018esrgan}. The adversarial loss $L_G^a$ is a symmetrical form with discriminator loss (Eq.~\ref{eq:discriminator loss}) as:
\begin{align}
  L_G^a =& \mathbb{E}_{\mathbb{P}}[L_{BCE}(b,D_{\Theta_2}(X_{LR},X_{GLR}))]+\notag\\
      &\mathbb{E}_{\mathbb{Q}}[L_{BCE}(a,D_{\Theta_3}(X_{GLR},X_{LR}))].
\end{align}
Taking the  adversarial training, the parameters of DD-GAN generator $G_{\Theta_2}$ is optimized by generator loss $L_G$ as:
\begin{align}
  L_G = L_{con} + \lambda L_G^a
  \label{eq:17}
\end{align}
where $\lambda$ is the the coefficient to balance two loss terms. It should be noted that the follow-up Laplacian loss is not involved for DD-GAN, because the synthetic noises should be retained to get more degraded LR images.

\subsubsection{DuRCAN Restoration Loss Function}
Previous CNN-based SISR models~\cite{dong2015image,lim2017enhanced,zhang2018image} commonly utilize the pixel-oriented loss. Considering non-uniform noises greatly pollute the low-frequency area and edges of Cam-ScreenSR images are degraded, we add the Laplacian loss $L_{lap}$, which is inspired by the Laplace operation in image processing~\cite{rosenfeld1976digital} to sharpen the high-frequency edge and smooth the noises in low-frequency area. The Laplacian loss is defined on the 2D Laplace operator to minimize the $L_1$ distance between the filtering images of generated image $S_{\Theta_3}(X)$
and ground truth $Y$. Hence, the restoration loss function $L_{SR}(\cdot)$ is a weighted combination of
$L_1$ loss and $L_{lap}$ loss as:
\begin{align}
  L_{SR}&=L_1+\eta L_{lap}\notag\\&=L_1+\eta\frac{1}{w_lh_l}\sum_{i=1}^{w_l}\sum_{j=1}^{h_l}|\kappa_{lap}(Y)_{i,j}-\kappa_{lap}(S_{\Theta_3}(X))_{i,j}|,
  \label{6}
\end{align}
where $L_1$ loss is the main part and $\eta$ is the coefficient to balance two loss terms. $\kappa_{lap}(\cdot)$ denotes the filter with second order differential Laplace kernel~\cite{rosenfeld1976digital}, $w_l,h_l$ represent the size of filtering image. The effectiveness of Laplacian loss is demonstrated in Section~\ref{abla}.

\begin{table*}[t]
  \begin{center}
  \centering
  \caption{Average PSNR and SSIM results on three camera-screen degraded testing datasets with scale factor 4.  \textbf{Text} and \underline{text} indicate the best and the second best performance. }
  \label{table1}
  \centering
  \begin{tabular}{c|c|c|cc|cc|cc}
  \hline
  \hline
  \multirow{3}{*}{Algorithm}&\multirow{3}{*}{Scale}&\multirow{3}{*}{Parameters}&\multicolumn{2}{c}{Testing Set 1}&\multicolumn{2}{c}{Testing Set 2}&\multicolumn{2}{c}{Testing Set 3}\\
  &&&\multicolumn{2}{c}{Samsung S27R350 + Canon 760D}&\multicolumn{2}{c}{Lenovo X1 + Huawei P30}&\multicolumn{2}{c}{Dell IN2020M + iPhone 11}\\
  &&&\multicolumn{2}{c}{PSNR/SSIM}&\multicolumn{2}{c}{PSNR/SSIM}&\multicolumn{2}{c}{PSNR/SSIM}\\
  \hline
  Bicubic& $X4$&-&\multicolumn{2}{c}{16.24 / 0.6401}&\multicolumn{2}{c}{16.82 / 0.6622}&\multicolumn{2}{c}{19.31 / 0.6904} \\
  \hline
  SRCNN~\cite{dong2015image}& $X4$&15K&\multicolumn{2}{c}{18.55 / 0.6048 }&\multicolumn{2}{c}{18.37 / 0.6926}&\multicolumn{2}{c}{13.43 / 0.5615} \\
  VDSR~\cite{kim2016accurate}& $X4$&665K&\multicolumn{2}{c}{20.51 / 0.7025 }&\multicolumn{2}{c}{19.76 / 0.6983}&\multicolumn{2}{c}{13.96 / 0.5842} \\
  EDSR~\cite{lim2017enhanced}& $X4$&43090K&\multicolumn{2}{c}{21.93 / 0.7086 }&\multicolumn{2}{c}{21.32 / 0.7054}&\multicolumn{2}{c}{14.19 / 0.5933} \\
  ESRGAN~\cite{wang2018esrgan}& $X4$&17000K &\multicolumn{2}{c}{23.68 / 0.7140}&\multicolumn{2}{c}{23.91 / 0.7122}&\multicolumn{2}{c}{14.53 / 0.6014} \\
  RCAN~\cite{zhang2018image}& $X4$&15592K&\multicolumn{2}{c}{\underline{23.87} / \underline{0.7155}}&\multicolumn{2}{c}{23.33 / 0.7116}&\multicolumn{2}{c}{14.44 / 0.6045 } \\
  DuRCAN (Ours)& $X4$&5453K&\multicolumn{2}{c}{24.34 / 0.7224}&\multicolumn{2}{c}{24.07 / 0.7181}&\multicolumn{2}{c}{14.98 / 0.6178} \\
  \hline
  SRCNN~\cite{dong2015image} + DD-GAN& $X4$&5352K&\multicolumn{2}{c}{17.83 / 0.6858 }&\multicolumn{2}{c}{18.62 / 0.6922}&\multicolumn{2}{c}{19.84 / 0.6909} \\
  VDSR~\cite{kim2016accurate} + DD-GAN& $X4$&6002K&\multicolumn{2}{c}{19.97 / 0.7044}&\multicolumn{2}{c}{ 21.82 / 0.7068}&\multicolumn{2}{c}{20.63 / 0.7030} \\
  EDSR~\cite{lim2017enhanced} + DD-GAN& $X4$&49077K&\multicolumn{2}{c}{21.29 / 0.7051}&\multicolumn{2}{c}{23.48 / 0.7125}&\multicolumn{2}{c}{21.13 / 0.7061} \\
  ESRGAN~\cite{wang2018esrgan} + DD-GAN& $X4$& 22987K &\multicolumn{2}{c}{23.74 / 0.7120}&\multicolumn{2}{c}{\underline{24.20} / \underline{0.7207}}&\multicolumn{2}{c}{21.50 / 0.7033} \\
  RCAN~\cite{zhang2018image} + DD-GAN& $X4$&21580K&\multicolumn{2}{c}{23.81 / 0.7148}&\multicolumn{2}{c}{24.03 / 0.7179}&\multicolumn{2}{c}{\underline{21.86} / \underline{0.7075}} \\
  DuRCAN + DD-GAN(Final,Ours)& $X4$&11440K&\multicolumn{2}{c}{\textbf{24.82} / \textbf{0.7271} }&\multicolumn{2}{c}{\textbf{24.51} / \textbf{0.7240}}&\multicolumn{2}{c}{\textbf{22.19} /  \textbf{0.7103} } \\
  \hline
  \hline
  \end{tabular}
\end{center}
\end{table*}

\section{Experiments}
\label{exp}
For easier sorting through detailed results, we conduct comparison experiments and ablation analysis. We first introduce the datasets involved for our comparison experiments in Section~\ref{datasets} and our training setup in Section~\ref{setup}. Then, we train our model and several state-of-the-art methods (SOTA) on the Cam-ScreenSR training set. The cross-camera-screen evaluations are conducted in Section~\ref{expsr} to prove the generalization and effectiveness of our solution. The comparisons between models with and without DD-GAN have also be conducted. Next, in Section~\ref{expbi}, we attempt to prove that camera-screen degradation can effectively improve the performance for typical SISR task. Hence, we finetune and evaluate those Cam-ScreenSR-trained models in typical bicubic interpolation (BI) datasets. Moreover, we conduct the qualitative evaluations on real-world photographs to compare our model against other SOTAs in Section~\ref{expre}. And the comparisons of computational cost is presented in Section~\ref{expco}. Finally, we conduct ablation studies in Section~\ref{setup} to clearly present the effects of dual residual blocks, residual channel attention blocks and Laplacian loss.

\subsection{Datasets}
\label{datasets}
We utilize the HR images of DIV2K~\cite{timofte2017ntire} as the ground truths and collect the corresponding LR images with camera-screen degradation for X4 SISR task. The original DIV2K is divided to the training (ID: 0001-0800) and testing sets (ID: 0801-0900). For training set, the LR image is captured with Samsung S27R350 + Canon 760D. And for more general validation, the testing sets has three versions with different camera-screen combinations, as shown in Table~\ref{Equipment}. To prove complicated camera-screen degradation is efficient for typical BI SISR task,
the camera-screen trained models, including our DuRCAN and other SOTAs, are finetuned with the original DIV2K training set~\cite{timofte2017ntire} and tested on popular BI datasets: Set5~\cite{bevilacqua2012low}, Set14~\cite{zeyde2010single}, BSD100~\cite{arbelaez2010contour}, Urban100~\cite{huang2015single}. Moreover, the real-captured photographs by Huawei P30 will be used to validate the generalization of our approach in real-world scene. Following previous works~\cite{dong2015image,kim2016accurate,kim2016deeply,tong2017image,lim2017enhanced,zhang2018image,ledig2017photo,wang2018esrgan}, the evaluation metrics in our work are PSNR and SSIM~\cite{wang2004image} indices. The SISR results are evaluated using the Y channel in the YCbCr space.

\subsection{Training Setup}
\label{setup}
Both the camera-screen SISR, typical BI SISR and ablation analysis apply the  setups in this section. The LR training images of Cam-ScreenSR and typical BI datasets are randomly cropped into $48\times 48$ with mini-batch size 16. The 800 training images are randomly rotated by $90^{\circ},180^{\circ},270^{\circ}$ and horizontally flipped for data augmentation.

To balance the distribution of LR inputs, we set the random rate between real LRs $X_{LR}$ and generated LRs $X_{GLR}$ with $4:1$, where the mixing rate in Eq.~\ref{eq3} is $\gamma=0.25$. The upper and lower limits of discriminator label smoothing in Eq.~\ref{eq:discriminator loss} are set as $\alpha = 0.2,\beta=0.8$. To balance different loss terms, the coefficients of generator and restoration loss functions are set as $\lambda=1\times 10^{-3}$ in Eq.~\ref{eq:17} and $\eta = 6 \times 10^{-3}$ in Eq.~\ref{6}.

We select several typical state-of-the-art (SOTA) SISR models for comparison, the opensource codes of which have been released, including: SRCNN~\cite{dong2015image}, VDSR~\cite{kim2016accurate}, EDSR~\cite{lim2017enhanced}, ESRGAN~\cite{wang2018esrgan} and RCAN~\cite{zhang2018image}. The proposed DD-GAN and DuRCAN are jointly trained and the SISR testing only uses the trained DuRCAN. The learning rate is fixed at $10^{-4}$ and halved every 50000 iterations. We use Adam~\cite{kingma2014adam} ($\beta_1=0.9,\beta_2=0.999$)
to optimize parameters of our network. All the experiments were conducted on NVIDIA Titan Xp GPUs.

\begin{figure*}[t]
  \centering
  \includegraphics[width = 18.5cm]{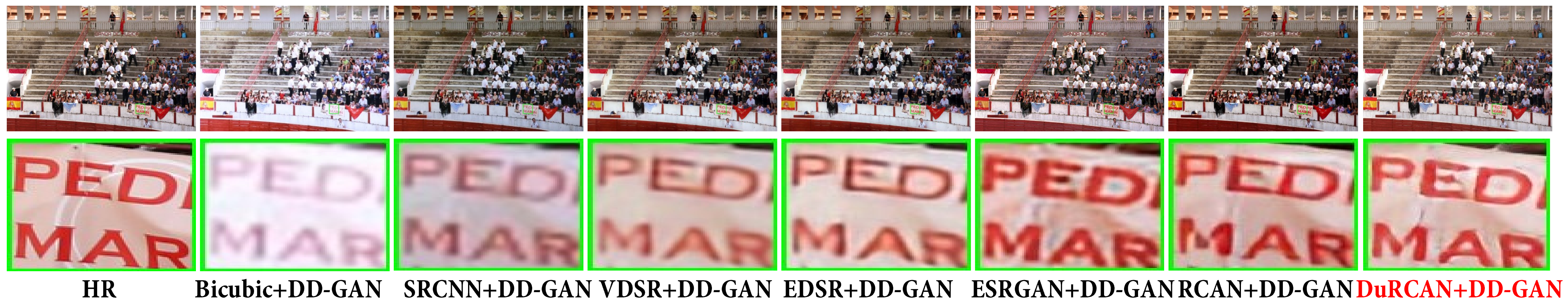}
  \caption{Visual comparisons of Img-0825 from the testing set 2, where upscaling factor is 4. We present 6 better performed models with DD-GAN data augmentation. The HR and Bicubic baselines are also shown in the first and second columns. Our joint model delivers best visual quality with sharper edge, less artifacts and more appropriate color enhancement.}
  \label{fig:screen_compare}
\end{figure*}
\subsection{SISR Models Trained on Camera-Screen Degradation}
\label{expsr}
\begin{figure}[t]
  \centering
  \includegraphics[width = 8.9cm]{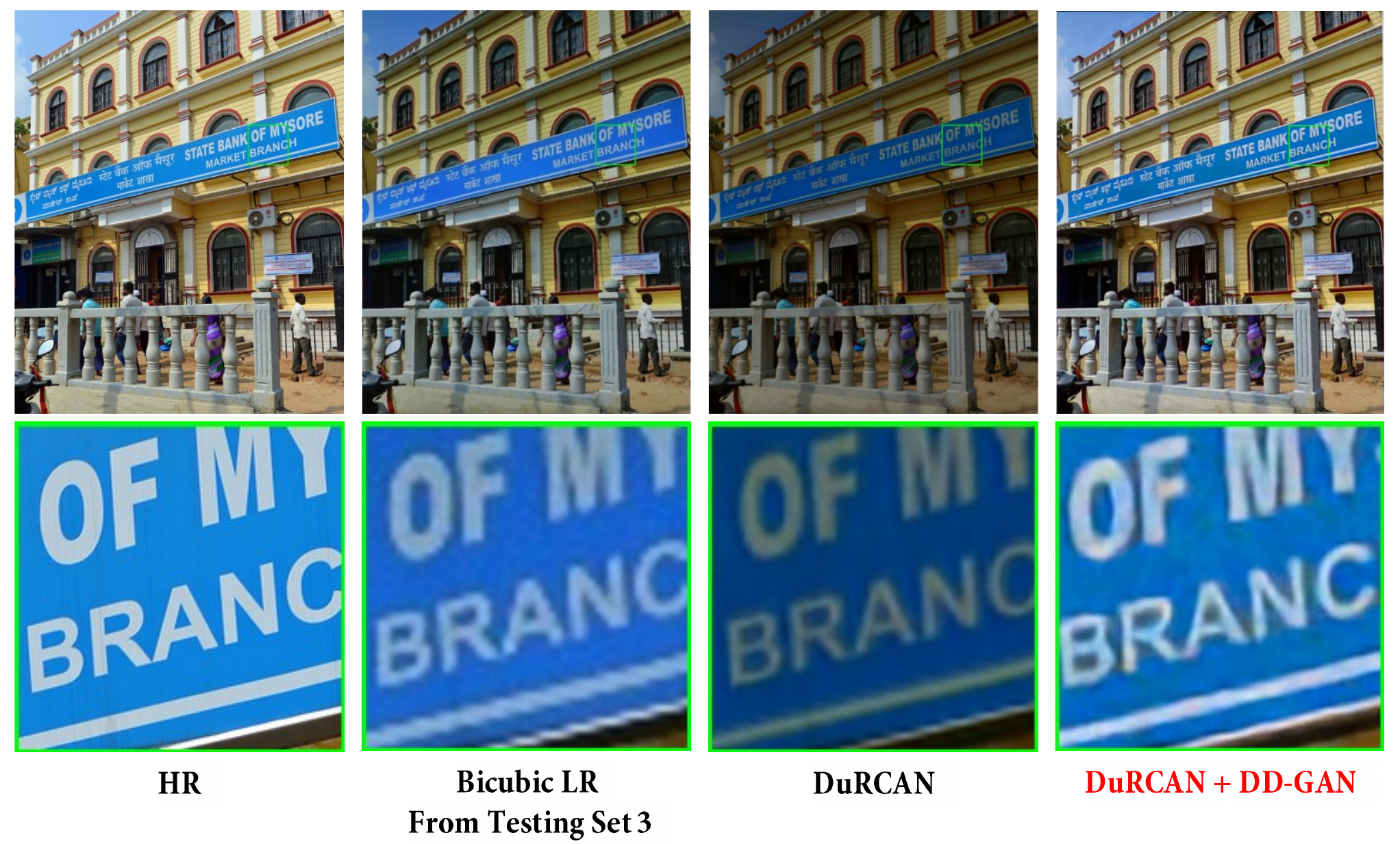}
  \caption{Visual comparisons (X4) on the testing set 3 with Dell IN2020M and iPhone 11. The first and second rows show the '0891' HR and bicubic baselines. The latter two rows show the results of DuRCAN separately trained with and without DD-GAN.}
  \label{fig:overfit_withoutGAN}
\end{figure}
As we mentioned before, what researchers usually call "super resolution" task is purely to restore an image that has been prefiltered by some kernel and then downsampled. But our work focuses on SISR with more complicated camera-screen degradation. To the best of authors' knowledge, there is no previous solution in this task to compare. Considering our model still attempt to restore images for higher resolution, we compare the proposed method with typical state-of-the-art (SOTA) SISR models including: SRCNN~\cite{dong2015image}, VDSR~\cite{kim2016accurate}, EDSR~\cite{lim2017enhanced}, ESRGAN~\cite{wang2018esrgan} and RCAN~\cite{zhang2018image}.
We trained those models on Cam-ScreenSR training dataset and tested them on three testing datasets with different camera-screen combinations to evaluate the robustness of SISR methods. Moreover, in order to validate that the generated LR images from DD-GAN is benefit for avoiding overfitting in specific degradation, we separately train the models with and without DD-GAN. The experiments of those existing methods are conducted with the released opensource codes.

\subsubsection{Quantitative Analyses}
As the quantitative results shown in Table~\ref{table1}, we can make some analyses as follows:

The 1st row in Table~\ref{table1} presents the bicubic baselines of three testing sets. The baseline of testing set 1 is lower than testing set 2 because the Lenovo X1 monitor with higher resolution provides clearer LR images. As Fig.~\ref{RGBdis} shows, the distributions of testing set 1 and 2 are relatively similar, which explains the slight gap between the baselines in those two testing sets. Both the RGB distributions of Fig.~\ref{fig:diff_screen} and visual examples in Fig.~\ref{RGBdis} reveal that the images of testing set 3 suffer less overexposure and color distortion. Hence, the third bicubic result is the higher than other two results.

The 2nd to 7th rows in Table~\ref{table1} show the results where SISR models are trained without DD-GAN. Our proposed DuRCAN outperforms the SOTA models. It is also clear that the PSNR/SSIM indexs of all the SISR models on testing set 1 has great improvements than bicubic baseline. Because the same camera-screen equipment is used, the training set and testing set 1 have the consistent degradation. However, learning the specific degradation pattern leads to the overfitting of CNN models, which results in the great performance deterioration on other degradated LR images.
The bicubic baselines reveal that the LR images of testing set 2 have higher quality than testing set 1. However, the SISR models trained without DD-GAN bring less improvements on testing set 2. When testing on testing set 3 with much different degradation pattern, all the CNN-based SISR models produce worst results, even lower than the bicubic baseline in the first row. Overfitting on specific degradation greatly deviates the learning representation of DNN models and limits the generalization  in real-world applications. When the variety of collected data is limited, exploring the effective data augmentation is necessary.

The 8th to 13th rows in Table~\ref{table1} show the results where SISR models are jointly trained with DD-GAN. Because of the big gap between the HRs and real camera-screen degraded images, the proposed DD-GAN can generate more various LRs, as shown in Fig.~\ref{fig:generateLR}. Except on testing set 3, the performance of most SOTA models + DD-GAN are slightly decreased, which means the generalization of those models is not enough to handle complicated degradation. After the DD-GAN generating more LR images with various degradation, the performance of our DuRCAN on both three testing sets get significant improvement than DD-GAN without DD-GAN and also achieves the best in Table~\ref{table1}. Specifically, the ESRGAN is not seriously influenced by data distribution, and utilizes the adversarial learning to enriche texture details by adding high-frequency noises. However, those uncontrollable noises greatly increase the training difficulty and pollute the output image, which leads to the less evaluation indexs compared with our DuRCAN and also limits the robustness in real-world images (seen in Section~\ref{expre}). The generated data enlarges the variety of image degradation and the dual residual convolution of our DuRCAN has great ability to handle those complicated degradations. Hence, the PSNR/SSIM growth of our model on testing set 3 is remarkable. Moreover, under the premise of better performance, the parameters of our DuRCAN is much less than existing SOTA models, like ESRGAN~\cite{wang2018esrgan} and RCAN~\cite{zhang2018image}, which proves the superiority of our model in real-world degradation.

\begin{table*}[htbp]
  \begin{center}
  \centering
  \caption{Average PSNR and SSIM results on four typical bicubic interpolation datasets with scale factor 4. The symbol "+" represents using the camera-screen pretrained model in Section~\ref{expsr}. \textbf{Text} and \underline{text} indicate the best and the second best performance.}
  \label{table2}
  \centering
  \begin{tabular}{c|c|c|c|c|c|c|c|c|c|c|c}
  \hline
  \hline
  \multirow{2}{*}{Model}&\multirow{2}{*}{Scale}&\multirow{2}{*}{Parameters / Size}&\multicolumn{2}{c}{ Set5~\cite{bevilacqua2012low}}&\multicolumn{2}{c}{ Set14~\cite{zeyde2010single}}&\multicolumn{2}{c}{BSD100~\cite{arbelaez2010contour}}&\multicolumn{2}{c}{Urban100~\cite{huang2015single}}\\
  &&&\multicolumn{2}{c}{PSNR / SSIM}&\multicolumn{2}{c}{PSNR / SSIM}&\multicolumn{2}{c}{PSNR / SSIM}&\multicolumn{2}{c}{PSNR / SSIM}\\
  \hline
  Bicubic& $X4$&-&\multicolumn{2}{c}{28.42 / 0.8104}&\multicolumn{2}{c}{26.00 / 0.7027}&\multicolumn{2}{c}{25.96 / 0.6675 }&\multicolumn{2}{c}{23.14 / 0.6577}\\
  \hline
  SRCNN~\cite{dong2015image}& $X4$&15K&\multicolumn{2}{c}{30.48 / 0.8628}&\multicolumn{2}{c}{27.50 / 0.7513}&\multicolumn{2}{c}{26.90 / 0.7101}&\multicolumn{2}{c}{24.52 / 0.7221}\\
  VDSR~\cite{kim2016accurate}& $X4$&665K&\multicolumn{2}{c}{31.35 / 0.8830}&\multicolumn{2}{c}{28.02 / 0.7680}&\multicolumn{2}{c}{27.29 / 0.7251} &\multicolumn{2}{c}{25.18 / 0.7524}\\
  EDSR~\cite{lim2017enhanced}& $X4$&43090K&\multicolumn{2}{c}{32.46 / 0.8968}&\multicolumn{2}{c}{28.80 / 0.7876}&\multicolumn{2}{c}{27.71 / 0.7420 }&\multicolumn{2}{c}{26.64 / 0.8033}\\
  ESRGAN~\cite{wang2018esrgan}& $X4$&17000K&\multicolumn{2}{c}{32.60 / 0.9002}&\multicolumn{2}{c}{28.88 / 0.7896}&\multicolumn{2}{c}{27.76 / 0.7432}&\multicolumn{2}{c}{26.73 / 0.8072}\\
  RCAN~\cite{zhang2018image}& $X4$&15592K&\multicolumn{2}{c}{32.63 / 0.9002}&\multicolumn{2}{c}{28.87 / 0.7889}&\multicolumn{2}{c}{ 27.77 / 0.7436 }&\multicolumn{2}{c}{26.82 / 0.8087}\\
  DuRCAN (Ours)& $X4$&5453K&\multicolumn{2}{c}{32.61 / 0.8996}&\multicolumn{2}{c}{28.85 / 0.7884}&\multicolumn{2}{c}{27.74 / 0.7429}&\multicolumn{2}{c}{26.84 / 0.8091}\\
  \hline
  SRCNN+~\cite{dong2015image}& $X4$&59KB&\multicolumn{2}{c}{30.50 / 0.8643}&\multicolumn{2}{c}{27.59 / 0.7518}&\multicolumn{2}{c}{26.96 / 0.7151}&\multicolumn{2}{c}{24.60 / 0.7233}\\
  VDSR+~\cite{kim2016accurate}& $X4$&2.55MB&\multicolumn{2}{c}{31.39 / 0.8835}&\multicolumn{2}{c}{28.10 / 0.7686}&\multicolumn{2}{c}{27.33 / 0.7261} &\multicolumn{2}{c}{25.27 / 0.7540}\\
  EDSR+~\cite{lim2017enhanced}& $X4$&164MB&\multicolumn{2}{c}{32.58 / 0.8984}&\multicolumn{2}{c}{28.84 / 0.7882}&\multicolumn{2}{c}{27.79 / 0.7431 }&\multicolumn{2}{c}{26.78 / 0.8073}\\
  ESRGAN+~\cite{wang2018esrgan}& $X4$&63.8MB&\multicolumn{2}{c}{\underline{32.63} / \underline{0.9005}}&\multicolumn{2}{c}{28.89 / 0.7894}&\multicolumn{2}{c}{\underline{27.80} / \underline{0.7433}}&\multicolumn{2}{c}{26.84 / 0.8081}\\
  RCAN+~\cite{zhang2018image}& $X4$&59.7MB&\multicolumn{2}{c}{\textbf{32.70} / \textbf{0.9007}}&\multicolumn{2}{c}{\underline{28.90} / \underline{0.7896}}&\multicolumn{2}{c}{ \textbf{27.81} / \textbf{0.7439} }&\multicolumn{2}{c}{\underline{26.87} / \underline{0.8099}}\\
  DuRCAN+ (Final,Ours)& $X4$&20.8MB&\multicolumn{2}{c}{32.60 / 0.8982}&\multicolumn{2}{c}{\textbf{28.93} / \textbf{0.7900}}&\multicolumn{2}{c}{27.64 / 0.7415}&\multicolumn{2}{c}{\textbf{26.92} / \textbf{0.8116}}\\
  \hline
  \hline
\end{tabular}
\end{center}
\end{table*}

\begin{figure*}[t]
  \centering
  \includegraphics[width = 18cm]{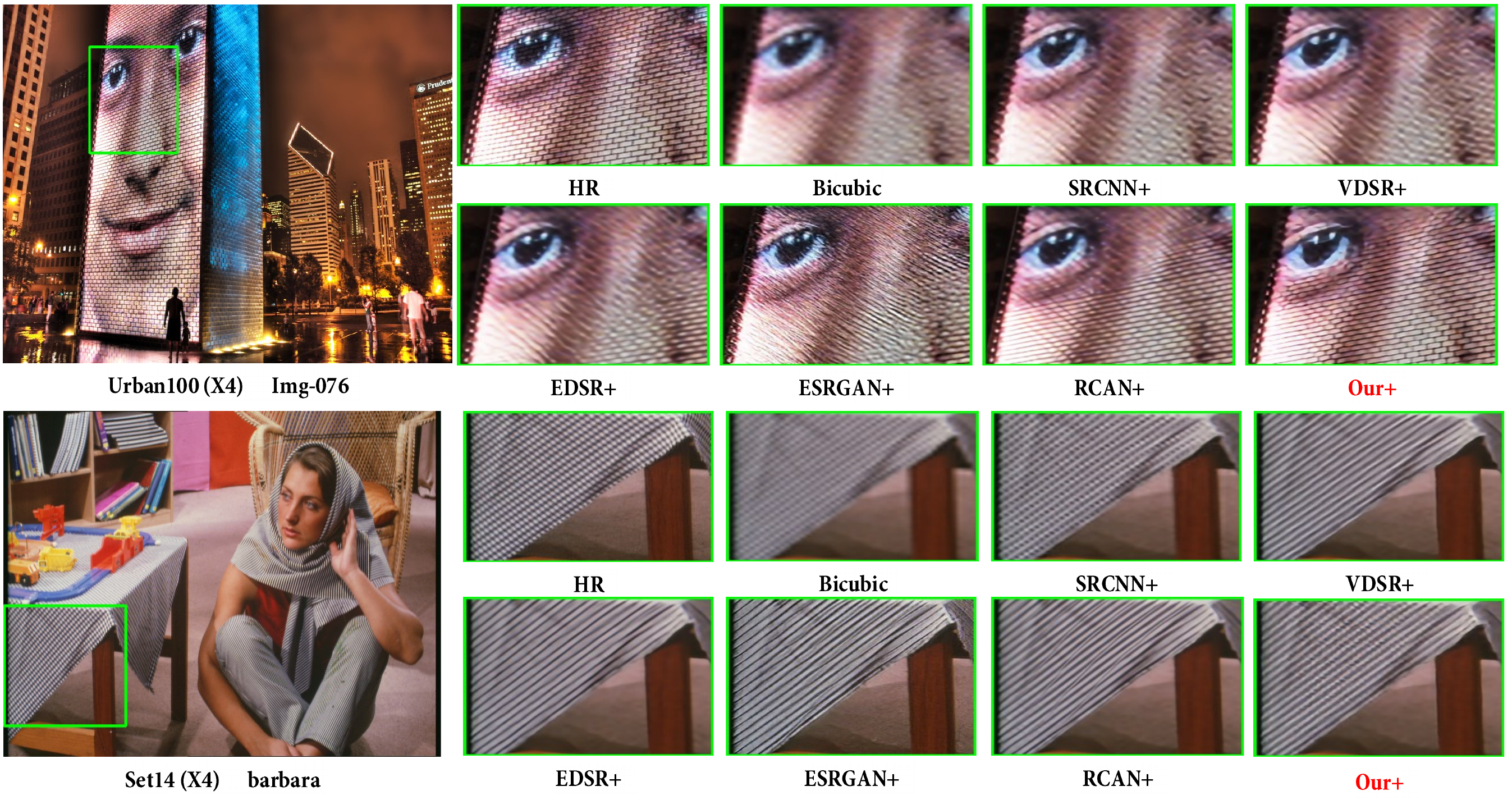}
  \caption{Examples of visual comparison (X4) on BI degradation. We present the results from finetuned models. Our DuRCAN restores more details, such as the tablecloth in Set14 "barbara" and the crosshatched pattern in Urban100 Img-076.}
  \label{fig:BI_compare}
\end{figure*}

\subsubsection{Qualitative Analyses}
 In Fig.~\ref{fig:screen_compare}, we show the bicubic baselines and visual comparisons of better SISR results with SISR models + DD-GAN on testing set 2 (Lenovo X1 + Huawei P30). Specifically, the visual comparisons of DuRCAN with and without DD-GAN on testing set 3 are illustrated in Fig.~\ref{fig:overfit_withoutGAN}, which validates the effectiveness of our proposed DD-GAN. It can be clearly seen that our model delivers best visual quality with sharper edge, less artifacts, especially appropriate color enhancement.

In this subsection, both the quantitative results and qualitative visual comparisons validate that: (1) the proposed DuRCAN has superiority to handle complicated real-world degradation; (2) the generative learning of proposed DD-GAN effectively enlarges the variety of degradations from limited real-captured data and the generated LR images from DD-GAN greatly enhance the robustness of SISR models.

\begin{figure*}[t]
  \centering
  \includegraphics[width = 18cm]{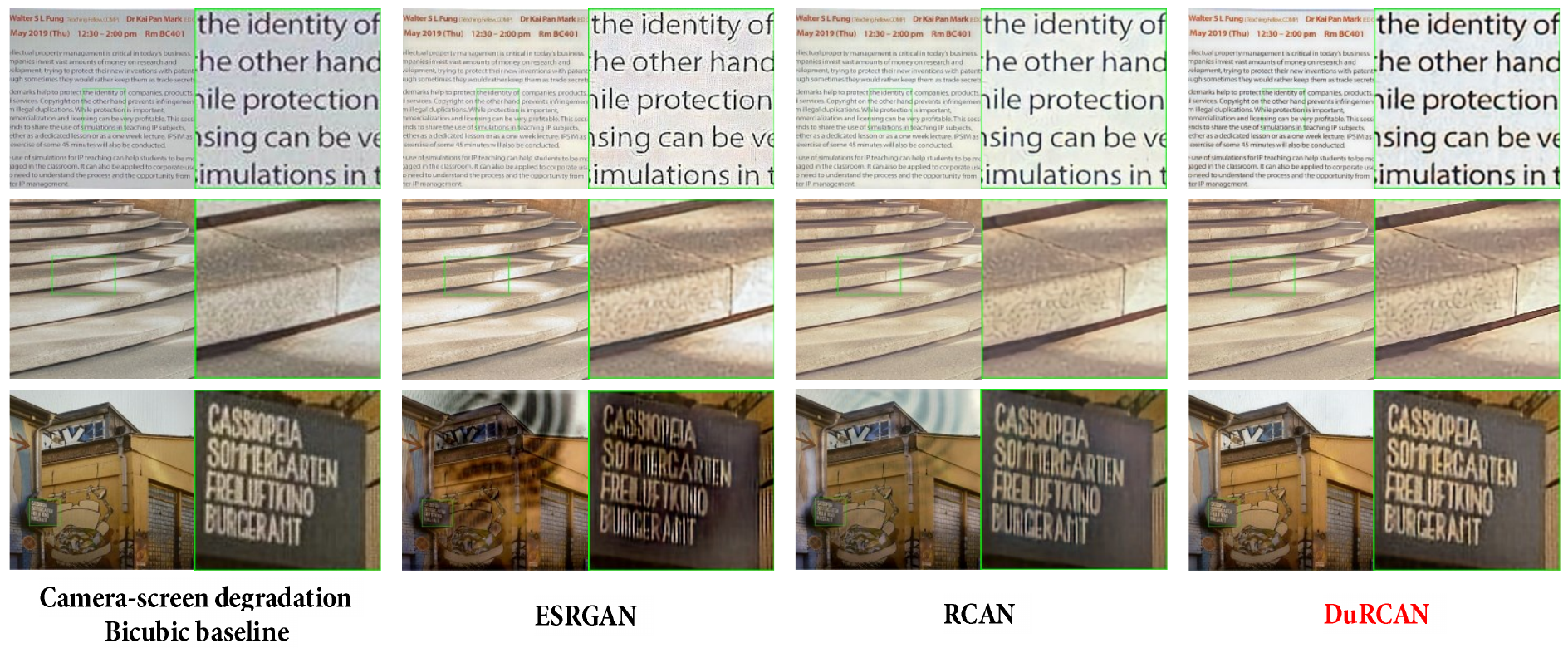}
  \caption{Examples of visual comparison (X4) on raw camera-screen degradation. The real-world images outside our dataset are not pre-processed. The RCAN and ESRGAN trained with Cam-ScreenSR are sensitive to the high-frequency Moir\'{e} pattern and can easily get corrupt by real-world disturbance. And our DuRCAN is more stable to produce the best results with less artifacts, sharper edges and better color enhancement.}
  \label{fig:sr_real}
\end{figure*}

\subsection{SOTA Comparisons on Typical Bicubic Datasets}
\label{expbi}
In Section~\ref{expsr}, we have proved that our proposed model has superiority over existing SISR models on real camera-screen degradation. It should be noticed that all those SOTAs were originally well-designed for typical "super resolution" task. Hence, we train all the methods on the original DIV2K dataset and evaluate them on four typical bicubic interpolation datasets to present a systemic comparison and verify the generalization of our joint model, including Set5~\cite{bevilacqua2012low}, Set14~\cite{zeyde2010single}, BSD100~\cite{arbelaez2010contour} and Urban100~\cite{huang2015single}. Besides directly citing the SISR results from the original papers of SRCNN~\cite{dong2015image}, VDSR~\cite{kim2016accurate}, EDSR~\cite{lim2017enhanced}, ESRGAN~\cite{wang2018esrgan} and RCAN~\cite{zhang2018image}, we utilize the pretrained camera-screen models in Section~\ref{expsr} to initialize the weights of parameters and finetune them on the typical BI DIV2K dataset.

\subsubsection{Quantitative Analyses}
As the quantitative results shown in Table~\ref{table2}, we can see that after using the pretrained weights from more complicated camera-screen degradation, the performance of all the models get improved (seen in the 8th to 13th rows of Table~\ref{table2}). This provides a new attempt for the improvement on SISR tasks that the SISR model can be appropriatly pretrained with complicated degradated images and then be finetuned in specific scene. Moreover, training with (DuRCAN+) and without (DuRCAN) pretrained initialization, our models both get competitive results among those well-designed SOTA methods for BI degradation. Although achieving a little bit lower performance on Set5 and BSD100 datasets, the DuRCAN outperforms on Set14 and Urban100 datasets. It is worth noting that the parameter quantities of the state-of-the-art perceptual-driven model ESRGAN and pixel-oriented model RCAN are 17000K and 15592K in x4 scale, while the parameter quantities of DuRCAN are only 5453K in x4 scale, which is about one-third (1/3) of those of ESRGAN and RCAN. With fewer parameters, our DuRCAN achieves the competitive and even better performance than those well-designed BI SISR models, which validates the effectiveness of our method.

\subsubsection{Qualitative Analyses}
In Fig.~\ref{fig:BI_compare}, we present the bicubic baselines and 6 better performed models with pretrained weights on two examples, including the Set14 "barbara" and Urban100 Img-076. Benefitting from the network structure and added Laplacian loss, many of the regular patterns that are undersampled and incorrectly reconstructed in the other methods are dealt with very well by our DuRCAN, such as the tablecloth in Set14 “barbara” and the crosshatched pattern in Urban100 Img-076. The Set14 and Urban100 datasets contain more regular graphics. The better quantitative and qualitative results on those two datasets validate that our method has superiority to enriche more details for regular graphics with sharper edges and less artifacts.

In this subsection, the experimental results validate that: (1) not only for complicated camera-screen degradation, the DuRCAN is also competitive with less parameters for typical bicubic interpolation SISR task; (2) for better performance, the SISR model can be appropriatly pretrained with complicated degradated images and then be finetuned in specific scene.

\begin{figure*}[t]
  \centering
  \includegraphics[width = 18cm]{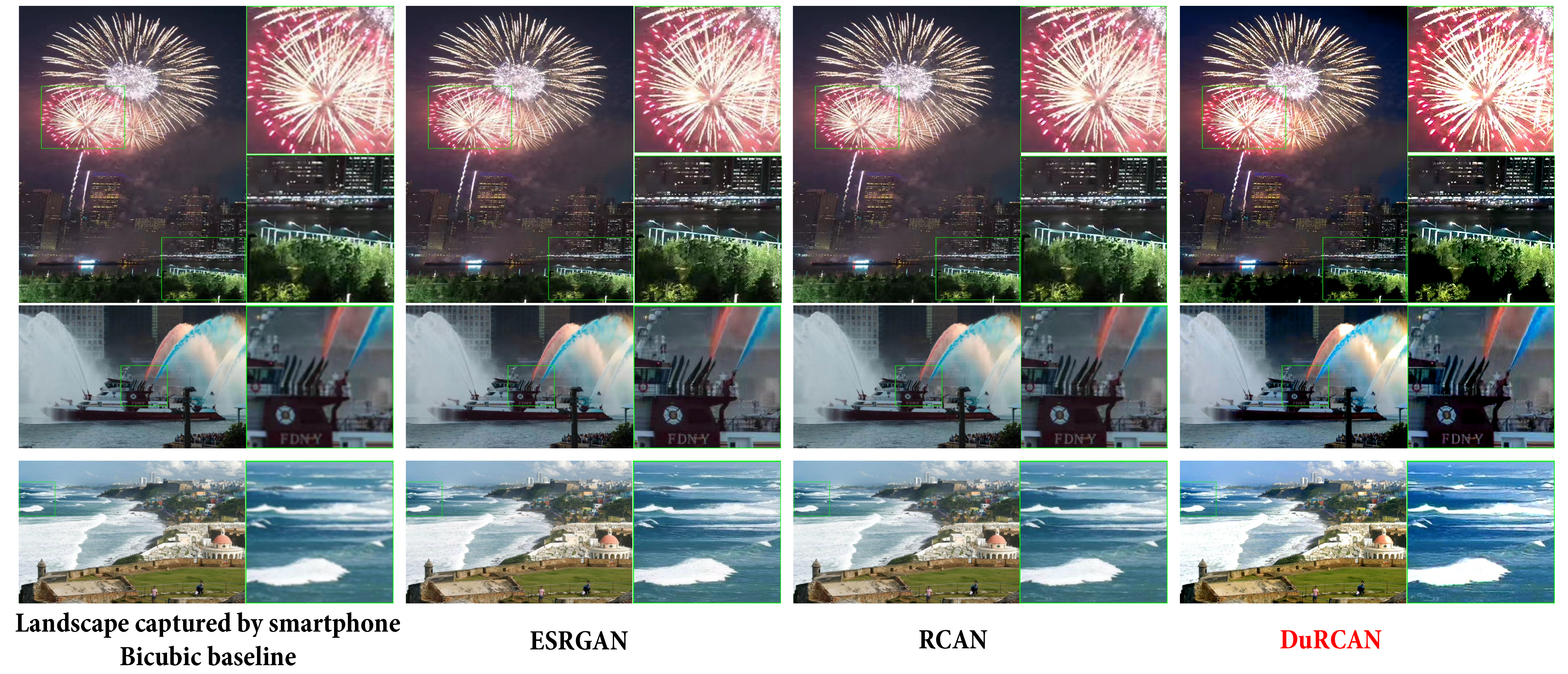}
  \caption{Examples of visual comparison for x4 SR images on landscape photographs captured by an iPhone 11 smartphone. We frozed the DuRBs and slightly finetuned the RCABs of Cam-ScreenSR trained models with BI images. Our DuRCAN delivers more comfortable visual results, especially with the excellent color enhancement.}
  \label{fig:ph_real}
\end{figure*}

\begin{table*}
  \begin{center}
  \centering
  \caption{Comparisons on PSNR/SSIM values, model  parameters (Pytorch-version) and average interface time. The testing set 1 (Samsung S27R350 + Canon 760D) with scale factor $\times 4$ is used for measurement. All the running time is calculated by a NVIDIA Titan XP GPU. }
  \label{co}
  \centering
  \begin{tabular}{c|c|c|c|c|c|c}
    \hline
    \hline
    &SRCNN~\cite{dong2015image}&VDSR~\cite{kim2016accurate}&EDSR~\cite{lim2017enhanced}&ESRGAN~\cite{wang2018esrgan}&RCAN~\cite{zhang2018image}&DuRCAN\\
    \hline
    Para.&\textbf{15K}&665K&43090K&17000K&15592K&5453K\\
    \hline
    Sec.&4.073&\textbf{0.721}& 2.163& 2.575& 1.659& 1.071\\
    \hline
    PSNR&17.83&19.97&21.29&23.74&23.81&\textbf{24.82}\\
    SSIM&0.6858&0.7044&0.7051&23.81&0.7148&\textbf{0.7271}\\
    \hline
    \hline
  \end{tabular}
  \end{center}
\end{table*}

\subsection{Qualitative Evaluations on Real-world Photographs}
\label{expre}
To further validate the generalization capability, we compare our model against SOTA models on more general real-world scenes. Since there are no ground-truth for the real-captured image, we conduct the perceptual judgement. Two scenes, including raw camera-screen degradation and landscape photographs captured by smartphone, are presented as follows. All the photographs are the original versions without data rectification.

\subsubsection{Raw Camera-Screen Degradation}
To further validate the generalization capability of our Cam-ScreenSR dataset and proposed joint model, we should compare our model with other models on raw camera-screen degraded images outside our dataset. We randomly selected high-quality images from the Google search and displayed them on the Lenovo X1 laptop. The degraded images are captured by an iPhone 11 smartphone and are directly fed into the trained models in Section~\ref{expsr} without data rectification. We selected three better performed methods trained on Cam-ScreenSR dataset for visual comparison, including ESRGAN~\cite{wang2018esrgan}, RCAN~\cite{zhang2018image} and our proposed DuRCAN.

The visual examples of three models and bicubic baselines are presented in Fig.~\ref{fig:sr_real}. As the visual results show, the models trained with our Cam-ScreenSR have advantages to handle the noises, color distortion and blurs influenced by the screen and camera. Without the image rectification in Section~\ref{Data Acquisition}, including the alignment, interpolation and average with continuous shoots, the raw images are more degraded. The third row in Fig.~\ref{fig:sr_real} reveals the RCAN and ESRGAN trained with Cam-ScreenSR are sensitive to the high-frequency Moir\'{e} pattern and can easily get corrupt by real-world disturbance. Compared to previous well-designed SOTAs for BI degradation, our DuRCAN has great robustness to handle more complicated situations in real-world degradation.

\subsubsection{Landscape Photographs Captured by Smartphone}
The super-resolution is an useful application for mobile phones to provide more comfortable visual experience for customers.  To estimate the reliability and practicability of our method for real-captured images, we also evaluate our model on landscape photographs captured by Smartphone. In this scene, we captured real-world landscape photographs with an iPhone 11. We also compare the proposed DuRCAN with two better performed SOTA methods, including ESRGAN~\cite{wang2018esrgan} and RCAN~\cite{zhang2018image}. To control the color enhancement appropriatly, the Cam-ScreenSR trained DuRCAN was finetuned with BI degraded images slightly. Specifically, we froze the parameters of dual residual block and finetuned two residual channel attention blocks. The network training was early stopped after 100 iterations.

The visual examples of three models and bicubic baselines are presented in Fig.~\ref{fig:ph_real}. As the visual results show, not only recovering more details, our finetuned model also enriches the image color appropriatly. After finetuning the targeted residual channel attention blocks, our DuRCAN can conduct color enhancement to produce buler sea and greener grass for example, while keeping the restoration ability. With the premise of recovering sufficient details in photographs, the appropriate color enhancement can provide more comfortable visual experience in real-world photography.

\subsection{Comparisons on Computational Cost}
\label{expco}
For fair comparison, we use the 6 Cam-ScreenSR trained models in Section~\ref{expsr}, including SRCNN~\cite{dong2015image}, VDSR~\cite{kim2016accurate}, EDSR~\cite{lim2017enhanced}, ESRGAN~\cite{wang2018esrgan}, RCAN~\cite{zhang2018image} and proposed DuRCAN, to evaluate the runtime on the computer with 2.2 GHz Intel i7 CPU and 1 NVIDIA Titan Xp GPU. The PSNR/SSIM values, model parameters and average interface time on testing set 1 (Samsung S27R350 + Canon 760D) are listed in Table~\ref{co}. It's clear that SRCNN has fewest parameters but achieve worst reconstruction performance with much slower running speed compared with other methods. Although VDSR recovers SR images with the fastest speed, this method still produces worse SR results than complicated models with more parameters. The proposed DuRCAN can achieve superior PSNR/SSIM values with faster reconstruction speed than ESRGAN and RCAN.

\subsection{Ablation Study}
\label{abla}
As discussed in Section~\ref{proposed network}, our joint solution contains four main components, including residual channel attention blocks (RCAB), dual residual blocks (DuRB), the added Laplacian loss and the data augmentation with downsampling degradation GAN (DD-GAN). We have validated the effectiveness of DD-GAN in Section~\ref{expsr}. Therefore, we will conduct ablation experiments of the rest three components as follows.

\subsubsection{Residual Channel Attention Blocks (RCAB)}
\label{abexpca}
To analyse the effects of residual channel attention machanism in our DuRCAN, firstly, we compare the quantitative performance between the DuRCANs with and without RCABs. The RCABs removed DuRCAN (\emph{Base}) and intact DuRCAN were trained on camera-screen degraded training set and evaluated on three testing sets with DD-GAN.

The quantitative results are listed in Table~\ref{RCAB removed}. It can be seen that the quantitative results of \emph{Base} model and DuRCAN are similar in testing set 1. In testing set 2 and 3, the intact DuRCAN outperforms \emph{Base} model, which verifies the effectiveness of RCABs. As we have described in Section~\ref{Data Acquisition}, the LR images of training set and testing set 1 are degraded with same equipment. But the degradation patterns of testing set 2 and 3 are different from training set. It means that after involving the residual channel attention machanism, the generalization of our model with different camera-screen combinations is greatly enhanced.

\begin{table}[t]
  \begin{center}
  \centering
  \caption{SISR results of models with (DuRCAN) and without (\emph{Base}) residual channel attention blocks on three camera-screen degraded testing datasets. The DD-GAN data augmentation is also jointly applied. }
  \label{RCAB removed}
  \centering
  \begin{tabular}{c|c|c|c}
  \hline
  \hline
  Cam-ScreenSR&\multirow{2}{*}{Scale}&\emph{Base}&\emph{Base} + RCABs (DuRCAN)\\
  Testing Set&&  PSNR / SSIM&  PSNR / SSIM\\
  \hline
  1&X4 &24.78 / 0.7265&\textbf{24.82} / \textbf{0.7271}\\ 2&X4 &23.82 / 0.7146&\textbf{24.51} / \textbf{0.7240}\\
  3&X4 &21.47 / 0.7013 &\textbf{22.19} /  \textbf{0.7103}\\
  \hline
  \hline
  \end{tabular}
\end{center}
\end{table}

\begin{figure}[t]
  \centering
  \includegraphics[width = 8.9cm]{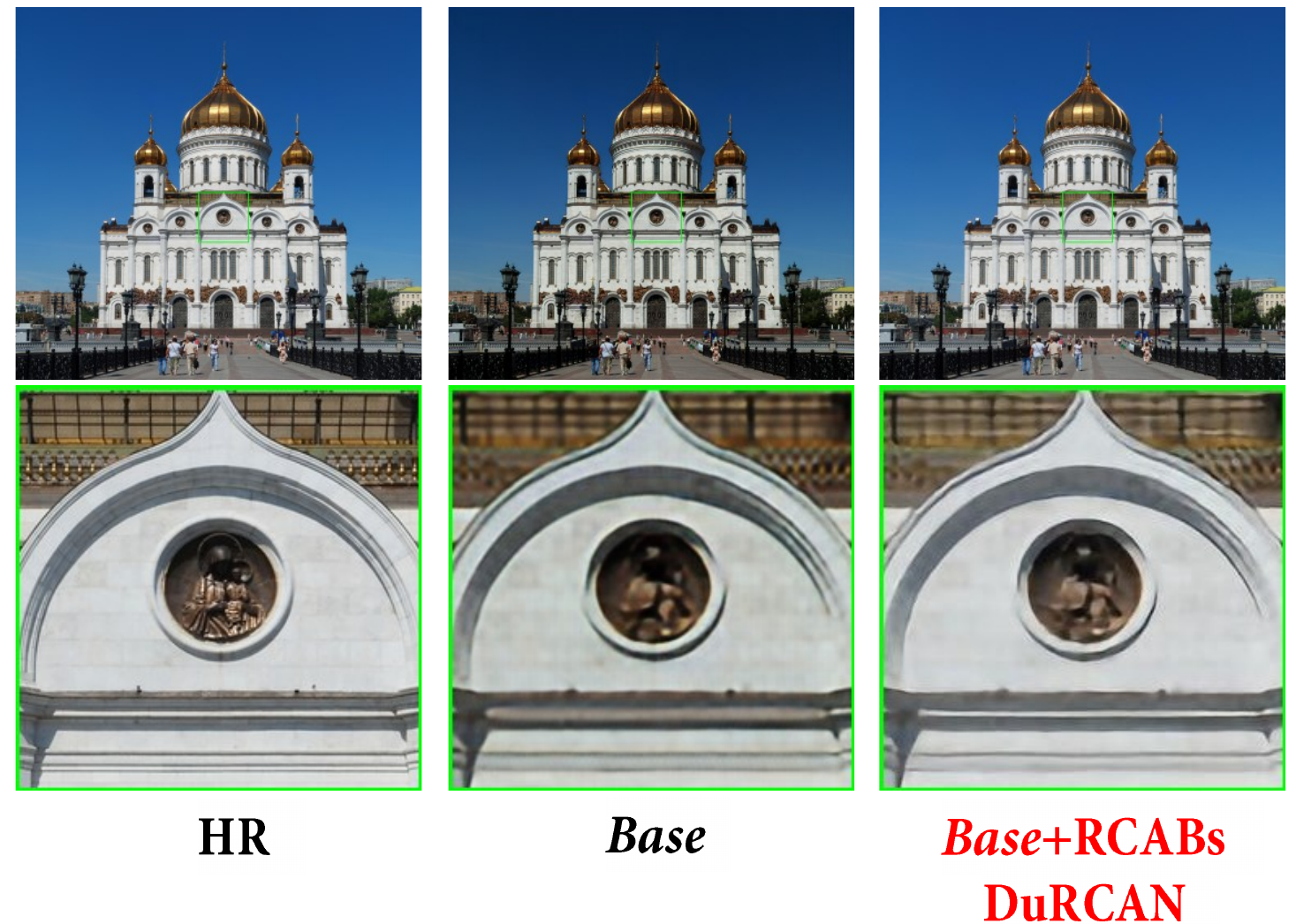}
  \caption{After removing RCAB, the SISR result of DIV2K Img-0879 produced by \emph{Base} model has acceptable noises and artifacts, but it has poor color distortion, which means RCABs focus more on the color calibration.}
  \label{fig:effect_RCAB}
\end{figure}

Moreover, the visual evaluation can clearly present the effects of RCABs. Specifically, we remove the RCABs from the intact trained DuRCAN and list the visual results in Fig.~\ref{RCAB removed}. The RCAB-removed result has acceptable noises and artifacts, but it has poor color distortion compared with HR image and SR result of DuRCAN, which reveals the color adjustment ability of \emph{Base} model is greatly weakened. Therefore, the residual channel attention blocks of our model have the limited influence on the SR definition, but focus more on the color calibration. It also provides us a way to specifically finetune the RCABs to control the color enhancement ability of our model.

\subsubsection{Dual Residual Blocks (DuRB)}
\label{abexpdu}
\begin{table}[t]
  \begin{center}
  \centering
  \caption{SISR results of DuRCAN with different DuRB configurations on three Cam-ScreenSR testing datasets. The DD-GAN data augmentation is also jointly applied.}
  \label{DuRB ablation}
  \centering
  \begin{tabular}{c|c|c|c}
  \hline
  \hline
  Model&Para.&Sec.&Dual kernel size\\
  \hline
  \multirow{3}{*}{DuRCAN-6\_s}&\multirow{3}{*}{1978K}&\multirow{3}{*}{0.4756}& $[3,3],[5,3],$\\
  &&&$[7,5],[7,5],$\\
  &&&$[7,3],[5,3]$\\
  \hline
  \multirow{3}{*}{DuRCAN-6}&\multirow{3}{*}{3518K}&\multirow{3}{*}{{0.6996}}& $[5,3],[7,5],$\\
  &&&$[11,7],[11,7],$\\
  &&&$[11,5],[7,5].$\\
  \hline
  \multirow{4}{*}{DuRCAN-12}&\multirow{4}{*}{5453K}&\multirow{4}{*}{1.071}& $[5,3],[5,3],[7,3],$\\
  &&&$[7,5],[11,5],[11,7],$\\
  &&&$[11,7],[11,5],[7,5],$\\
  &&&$[7,3],[5,3],[5,3].$\\
  \hline
  \multirow{6}{*}{DuRCAN-18}&\multirow{6}{*}{9878K}&\multirow{6}{*}{1.529}& $[5,3],[5,3],[5,3],$\\
  &&&$[7,5],[7,5],[7,5],$\\
  &&&$[11,7],[11,7],[11,7],$\\
  &&&$[11,7],[11,7],[11,7],$\\
  &&&$[11,5],[11,5],[11,5],$\\
  &&&$[7,5],[7,5],[7,5]$\\
  \hline
  \hline
  \multirow{2}{*}{Model}&\multirow{2}{*}{Scale}&Camera-Screen&\multirow{2}{*}{PSNR / SSIM}\\
  &&Testing Set&\\
  \hline
  \multirow{3}{*}{DuRCAN-6\_s}&\multirow{3}{*}{X4}&1& 23.89 / 0.7129 \\
  &&2& 23.95 / 0.7163 \\
  &&3& 20.89 / 0.6822\\
  \hline
  \multirow{3}{*}{DuRCAN-6}&\multirow{3}{*}{X4}&1&24.21 / 0.7205 \\
  &&2& 24.26 / 0.7207\\
  &&3& 21.45 / 0.6984\\
  \hline
  \multirow{3}{*}{DuRCAN-12}&\multirow{3}{*}{X4}&1&\underline{24.82} / \underline{0.7271} \\
  &&2&\underline{24.51} / \underline{0.7240} \\
  &&3&\underline{22.19} / \underline{0.7103} \\
  \hline
  \multirow{3}{*}{DuRCAN-18}&\multirow{3}{*}{X4}&1& \textbf{24.84 / 0.7275}\\

  &&2&\textbf{24.56 / 0.7242} \\

  &&3&\textbf{22.21 / 0.7104}  \\
  \hline
  \hline
  \end{tabular}
\end{center}
\end{table}

For SISR task, receptive field determines whether the ability of model is good enough to explore the relationships of neighbor pixels and recover the missing contextual information~\cite{kim2016deeply}. As the pooling operation will discard the image details, existing SISR models focus on increasing the network depth and enlarging the convolutional kernel size~\cite{dong2015image,kim2016accurate,kim2016deeply,tong2017image,lim2017enhanced,ledig2017photo,wang2018esrgan,zhang2018image}.
The configuration of dual residual blocks (DuRBs) determines the receptive field of our DuRCAN. Therefore, we conduct the ablation comparisons between different depth and kernel size settings. We changed the depth of DuRBs $d=6,12,18$, named as DuRCAN-6, DuRCAN-12 (our proposed model) and DuRCAN-18 to evaluate their performance respectively. And we also structured a DuRCAN-6 with smaller kernel sizes, named as DuRCAN-6\_s. The model configurations of dual kernel size are listed in Table.~\ref{DuRB ablation}. All those four models were evaluated in three camera-screen degraded testing sets.

The quantitative results are shown in Table.~\ref{DuRB ablation}. We can see that keeping the same depth of DuRBs, the DuRCAN-6 performs better than DuRCAN-6\_s in all three testing sets, which reveals the larger kernel size is effective for our DuRCAN structure. With different depth, the deeper DuRCAN-18 performed better than the shallow DuRCAN-6 and DuRCAN-12. It should be noticed that when increasing the depth of DuRBs, the performance of DuRCAN-18 gets marginal improvement than DuRCAN-12, but the model parameter and execution time are greatly increased. Considering that the DuRCAN are jointly trained with generative learning network DD-GAN, we choose the configuration of DuRCAN-12 as our proposed method to balance the computing cost and SISR performance.

\begin{figure}[t]
  \centering
  \includegraphics[width = 8.9cm]{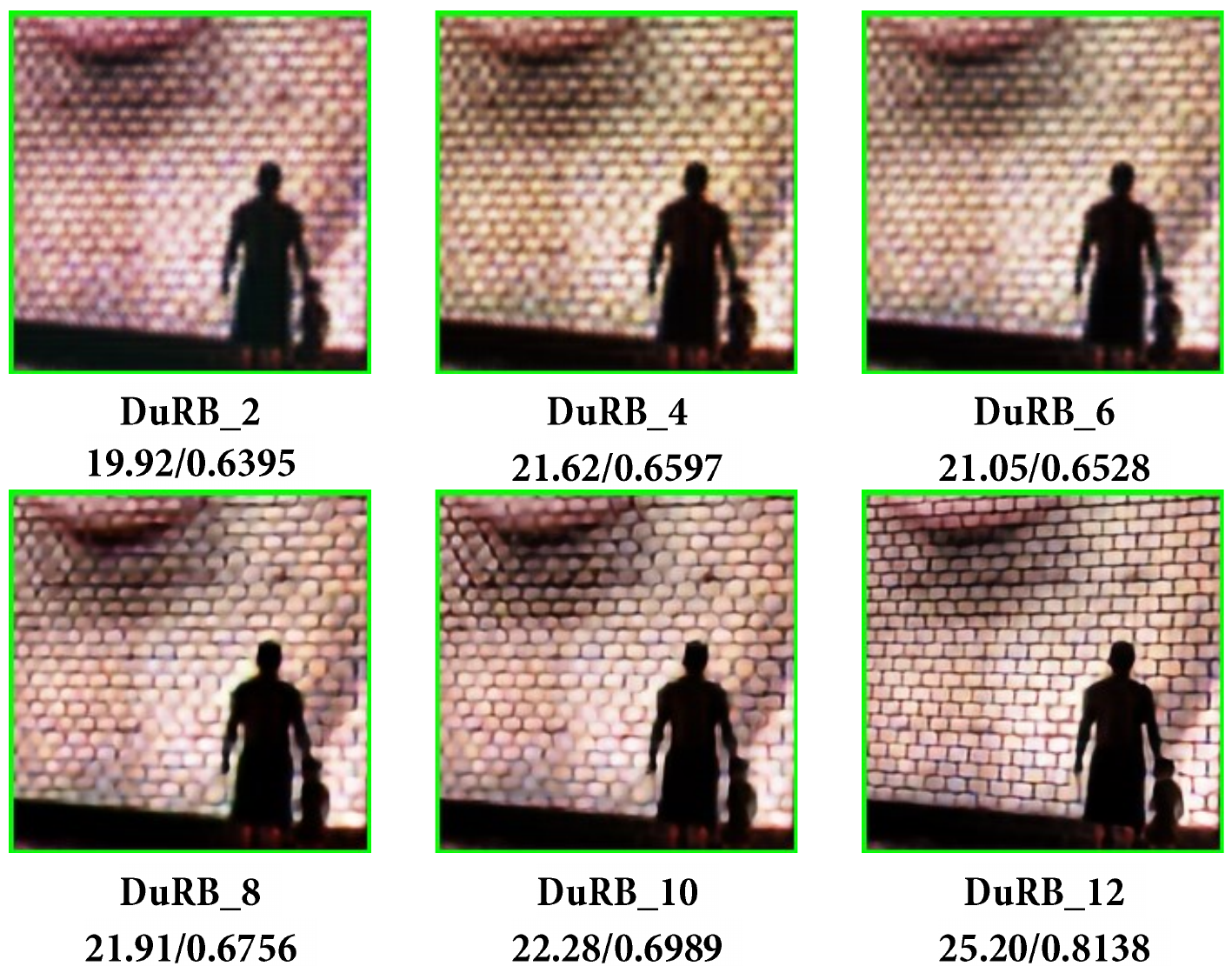}
  \caption{As the depth of DuRB increases, the X4 SR results of Urban100 Img-076 broadly become clearer and contain more textures, which reveals the dual residual operations pay attention on restoring the details.}
  \label{fig:effect_DuRB}
\end{figure}

Moreover, we visualize the recovering features of the DuRB to verify the effects of stacked DuRBs. In order not to involve extreme color distortion, the BI trained DuRCAN in Section~\ref{expbi} is applied. We recover the output features from every two layers of DuRBs to get the SR results. As Fig.~\ref{fig:effect_DuRB} shows, the deeper DuRB produces the better SR result. Combining the previous ablation experiment of residual channel attention machanism, it can be seen that the dual residual operations pay more attention on restoring the details from complicated degradations. Without the limitations of computing resources and execution time, the deeper DuRG can be involved to improve the SR definition.

\begin{table}[t]
  \begin{center}
  \centering
  \caption{SISR results of DuRCAN with different loss functions on Cam-ScreenSR testing sets. The DD-GAN data augmentation is also jointly applied.}
  \label{Laplacian loss SR}
  \centering
  \begin{tabular}{c|c|c|c}
  \hline
  \hline
  Cam-ScreenSR&\multirow{2}{*}{Scale}&DuRCAN + $L_1$&DuRCAN + $L_{SR}$\\
  Testing Set&&  PSNR / SSIM&  PSNR / SSIM\\
  \hline
  1&X4 &24.03 / 0.7191&\textbf{24.82} / \textbf{0.7271}\\ 2&X4 &23.84 / 0.7123&\textbf{24.51} / \textbf{0.7240}\\
  3&X4 &20.95 / 0.7028&\textbf{22.19} /  \textbf{0.7103}\\
  \hline
  \hline
  \end{tabular}
\end{center}
\end{table}

\begin{table}[t]
  \begin{center}
  \centering
  \caption{SISR results of DuRCAN with different loss functions on typical BI testing sets. The DD-GAN data augmentation is also applied.}
  \label{Laplacian loss BI}
  \centering
  \begin{tabular}{c|c|c|c}
  \hline
  \hline
  Typical BI&\multirow{2}{*}{Scale}&DuRCAN + $L_1$&DuRCAN + $L_{SR}$\\
  Testing Set&&  PSNR / SSIM&  PSNR / SSIM\\
  \hline
  Set5~\cite{bevilacqua2012low}&X4 &32.57 / 0.8973&\textbf{32.60} / \textbf{0.8982}\\ Set14~\cite{zeyde2010single}&X4 &28.85 / 0.7890&\textbf{28.93} / \textbf{0.7900}\\
  BSD100~\cite{arbelaez2010contour}&X4 &27.53 / 0.7369&\textbf{27.64} /  \textbf{0.7415}\\
  Urban100~\cite{huang2015single}&X4&26.82 / 0.8077&\textbf{26.92} / \textbf{0.8116}\\
  \hline
  \hline
  \end{tabular}
\end{center}
\end{table}

\begin{figure}[t]
  \centering
  \includegraphics[width = 8.9cm]{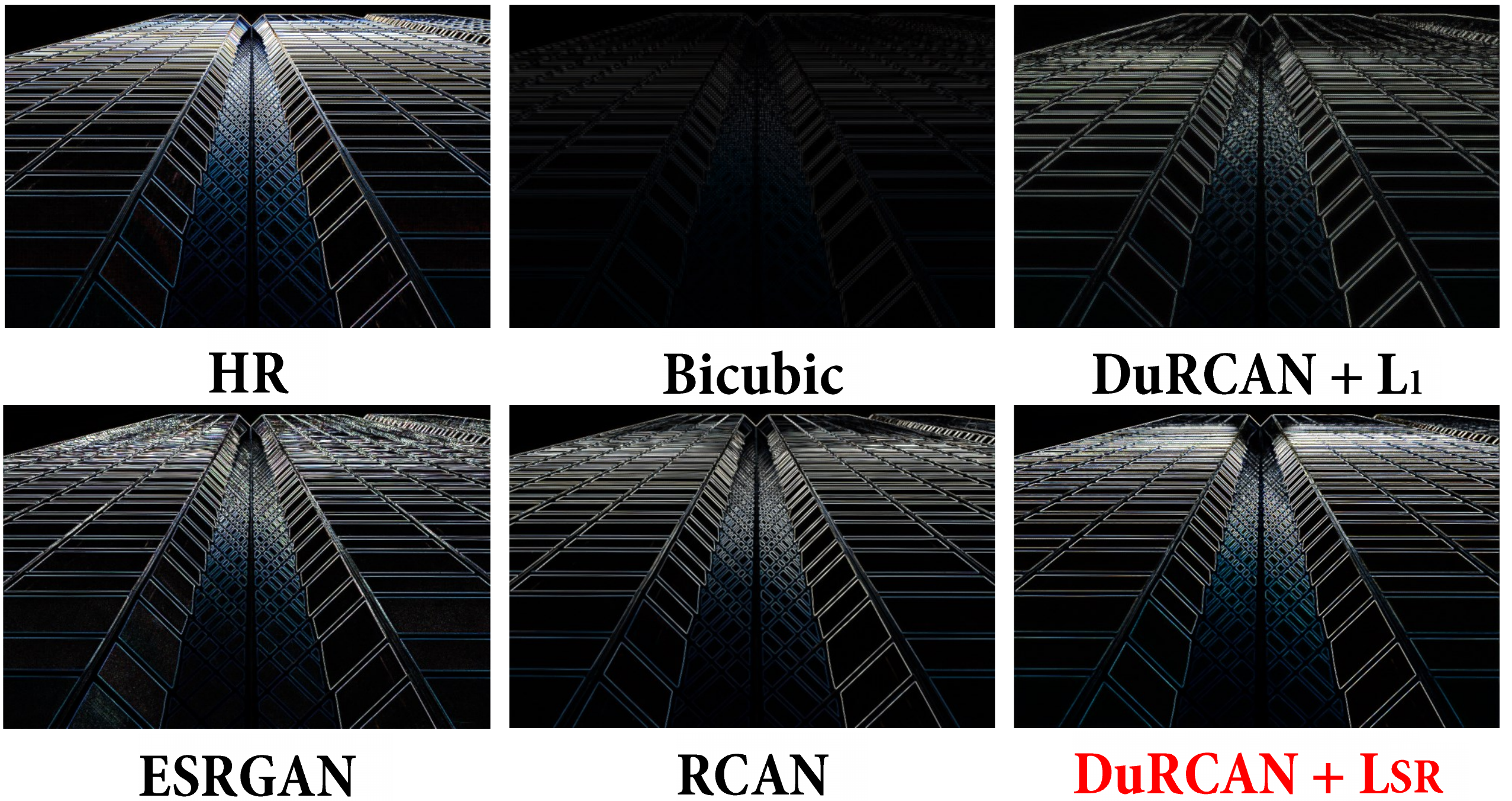}
  \caption{Visual comparisons of the edges processed by second order differential Laplacian operation. The DuRCAN trained with combined $L_{SR}$ loss can reconstruct sharper edges with more texture details and less noises.}
  \label{fig:Laplacian}
\end{figure}

\subsubsection{Laplacian Loss}
\label{abexpla}
It has been widely acknowledged that an image consists of the high-frequency and low-frequency messages. The low-frequency messages deliver the basic gray level of the image. And the high-frequency messages deliver the sharp changes of the pixel intensity, which reveal more details of the image~\cite{gonzalez2004digital}. Therefore, it's important to supervise our model to restore the high-frequency edges. Inspired by the second order differential Laplace operation commonly used in image processing~\cite{rosenfeld1976digital}, we involves the high-frequency supervised Laplacian loss function to minimize the $L_1$ distance between the HR and SR image processed by Laplacian operation. As Eq~\ref{6} presents, the Laplacian Loss $L_{lap}$ and $L_1$ loss are weighted combined as $L_{SR}$. To demonstrate the effectiveness of Laplacian loss, we compared the performance of the DuRCAN with $L_1$ and $L_{SR}$ on Cam-ScreenSR dataset respectively. The DD-GAN data augmentation was applied to correspond with the training method in Sec~\ref{expsr}. We also finetuned the models on typical BI DIV2K dataset and evaluated their performance on Set5~\cite{bevilacqua2012low}, Set14~\cite{zeyde2010single}, BSD100~\cite{arbelaez2010contour}, Urban100~\cite{huang2015single} datasets.

The quantitative results on camera-screen and typical BI degraded datasets are respectively shown in Table~\ref{Laplacian loss SR} and Table~\ref{Laplacian loss BI}. After adding the Laplacian loss, the performance of DuRCAN get improved on both Cam-ScreenSR and typical BI datasets. As visual examples shown in Fig~\ref{example}, the camera-screen degradation contains less high-frequency information than bicubic downsampling, because of much more blurs, noises and color distortion. Therefore, the quantitative improvements on Cam-ScreenSR dataset are greater than those on BI datasets after involving the high-frequency supervised loss $L_{lap}$.

Moreover, we visualize the edges of different methods processed by second order differential Laplacian operation in Fig.~\ref{fig:Laplacian}, including the bicubic baseline, ESRGAN~\cite{wang2018esrgan}, RCAN~\cite{zhang2018image} and our DuRCAN trained with $L_1$ loss and $L_{SR}$ loss. It can been seen that the bicubic SR image contains few high-frequency edges. The RCAN and DuRCAN trained with $L_1$ loss models outperform the bicubic baseline. Although ESRGAN enriches texture details by adding high-frequency noises, those uncontrollable noises severely limit the robustness in real-world images (seen in Fig.~\ref{fig:sr_real} and Fig.~\ref{fig:ph_real}). The DuRCAN trained with combined $L_{SR}$ loss can reconstruct sharper edges with more texture details. All the qualitative and quantitative results prove that the added Laplacian loss can efficiently supervise our proposed joint model to smooth the noises and sharpen the edges of SISR results.

\section{Discussion And Future Work}
Restoring the low-resolution images from the camera-screen degradation is a more sophisticated task. It's different from the pure "super-resolution", where the images are prefiltered by downsampling kernels. The camera-screen SISR task has to jointly solve problems such as denoising, sharpening the edge, fixing color distortion, etc.. Although those degradations have a long research history in image processing, they are inevitable for real-world SISR application. We have proved that involving them is benefit for the performance of SISR models. As our model focuses on SISR tasks, it's normal to choose previous SISR models for comparisons and we also conduct fair experiments on the same datasets to validate the effectiveness of our model. We believe that the joint solution for camera-screen SISR scene is valuable and should be encouraged, because the user can directly get the high-resolution output without multi-step processing.

In different real-world environments, there exists more extrame camera-screen degradation. In order not to decrease the stability of the proposed joint model by more extrame degradation, our data acquisition strategy partly simplifies the degradation: the repeated photo capturing partly weakens the stroboscopic effect; the image rectification and multiple image averaging partly weaken the noises, such as Moir\'{e} patterns; the monitors are set to the Standard mode to not cause extrame color distortion, and etc.. In the future work, more camera-screen combinations, more extrame degradation can be involve to expand our dataset and explore the novel model for better generalization. Moreover, although our Cam-ScreenSR dataset and joint model focus on the SISR task, they can be used in other image restoration tasks, such as image denoising, deblurring and color correction. We also leave it as our future work.

\section{Conclusion}
There exists a long standing problem that the SISR model trained with synthetic degradation has poor generalization on real-world image. In this paper, we made an first attempt to involve the degradation of camera-screen device for SISR task and proposed a data acquisition strategy to establish a baseline, Cam-ScreenSR dataset. Although the camera-screen degradation is complicated, our proposed joint model has the great ability to handle those degradations, which produces better visual results than previous SOTA models with sharper edge, less artifacts and appropriate color enhancement. We believe that our decent SISR solution will provide clearer visual experience for general users when they use their mobile phones to record contents on screens for convenience, simplicity and efficiency. Meanwihle, the appropriate color enhancement can also be accomplished, which is more easily perceived by human eyes.

\section*{Acknowledgment}

The authors would like to thank...

\ifCLASSOPTIONcaptionsoff
  \newpage
\fi

\bibliographystyle{IEEEtran}
\bibliography{ref}

\begin{thebibliography}{10}
\providecommand{\url}[1]{#1}
\csname url@samestyle\endcsname
\providecommand{\newblock}{\relax}
\providecommand{\bibinfo}[2]{#2}
\providecommand{\BIBentrySTDinterwordspacing}{\spaceskip=0pt\relax}
\providecommand{\BIBentryALTinterwordstretchfactor}{4}
\providecommand{\BIBentryALTinterwordspacing}{\spaceskip=\fontdimen2\font plus
\BIBentryALTinterwordstretchfactor\fontdimen3\font minus
  \fontdimen4\font\relax}
\providecommand{\BIBforeignlanguage}[2]{{%
\expandafter\ifx\csname l@#1\endcsname\relax
\typeout{** WARNING: IEEEtran.bst: No hyphenation pattern has been}%
\typeout{** loaded for the language `#1'. Using the pattern for}%
\typeout{** the default language instead.}%
\else
\language=\csname l@#1\endcsname
\fi
#2}}
\providecommand{\BIBdecl}{\relax}
\BIBdecl

\bibitem{glasner2009super}
D.~Glasner, S.~Bagon, and M.~Irani, ``Super-resolution from a single image,''
  in \emph{2009 IEEE 12th international conference on computer vision}.\hskip
  1em plus 0.5em minus 0.4em\relax IEEE, pp. 349--356.

\bibitem{yang2012coupled}
J.~Yang, Z.~Wang, Z.~Lin, S.~Cohen, and T.~S. Huang, ``Coupled dictionary
  training for image super-resolution,'' \emph{IEEE Transactions on Image
  Processing}, vol.~21, no.~8, pp. 3467--3478, 2012.

\bibitem{5702359}
H.~{Demirel} and G.~{Anbarjafari}, ``Discrete wavelet transform-based satellite
  image resolution enhancement,'' \emph{IEEE Transactions on Geoscience and
  Remote Sensing}, vol.~49, no.~6, pp. 1997--2004, 2011.

\bibitem{gunturk2003eigenface-domain}
B.~K. Gunturk, A.~U. Batur, Y.~Altunbasak, M.~H. Hayes, and R.~M. Mersereau,
  ``Eigenface-domain super-resolution for face recognition,'' \emph{IEEE
  Transactions on Image Processing}, vol.~12, no.~5, pp. 597--606, 2003.

\bibitem{yang2010image}
J.~Yang, J.~Wright, T.~S. Huang, and Y.~Ma, ``Image super-resolution via sparse
  representation,'' \emph{IEEE transactions on image processing}, vol.~19,
  no.~11, pp. 2861--2873, 2010.

\bibitem{bertero1998introduction}
M.~Bertero and P.~Boccacci, \emph{Introduction to inverse problems in
  imaging}.\hskip 1em plus 0.5em minus 0.4em\relax CRC press, 1998.

\bibitem{kim2010single}
K.~I. Kim and Y.~Kwon, ``Single-image super-resolution using sparse regression
  and natural image prior,'' \emph{IEEE transactions on pattern analysis and
  machine intelligence}, vol.~32, no.~6, pp. 1127--1133, 2010.

\bibitem{freedman2011image}
G.~Freedman and R.~Fattal, ``Image and video upscaling from local
  self-examples,'' \emph{ACM Transactions on Graphics (TOG)}, vol.~30, no.~2,
  p.~12, 2011.

\bibitem{huang2015single}
J.-B. Huang, A.~Singh, and N.~Ahuja, ``Single image super-resolution from
  transformed self-exemplars,'' in \emph{Proceedings of the IEEE Conference on
  Computer Vision and Pattern Recognition}, 2015, pp. 5197--5206.

\bibitem{yang2019deep}
W.~Yang, X.~Zhang, Y.~Tian, W.~Wang, J.~Xue, and Q.~Liao, ``Deep learning for
  single image super-resolution: A brief review,'' \emph{IEEE Transactions on
  Multimedia}, vol.~21, no.~12, pp. 3106--3121, 2019.

\bibitem{keys1981cubic}
R.~Keys, ``Cubic convolution interpolation for digital image processing,''
  \emph{IEEE Transactions on Acoustics, Speech, and Signal Processing},
  vol.~29, no.~6, pp. 1153--1160, 1981.

\bibitem{keysLanczos}
C.~E. Duchon, ``Lanczos filtering in one and two dimensions,'' \emph{J. Appl.
  Meteorol.}, vol.~18, no.~8, pp. 1016--1022, 1979.

\bibitem{dai2009softcuts}
S.~Dai, M.~Han, W.~Xu, Y.~Wu, Y.~Gong, and A.~K. Katsaggelos, ``Softcuts: A
  soft edge smoothness prior for color image super-resolution,'' \emph{IEEE
  Transactions on Image Processing}, vol.~18, no.~5, pp. 969--981, 2009.

\bibitem{Single}
X.~a. Q.Yan, Y.Xu, ``Single image superresolution based on gradient profile
  sharpness,'' \emph{IEEE Transactions on Image Processing}, vol.~24, no.~10,
  pp. 3187--3202, 2015.

\bibitem{dong2015image}
C.~Dong, C.~C. Loy, K.~He, and X.~Tang, ``Image super-resolution using deep
  convolutional networks,'' \emph{IEEE transactions on pattern analysis and
  machine intelligence}, vol.~38, no.~2, pp. 295--307, 2015.

\bibitem{dong2016accelerating}
C.~Dong, C.~C. Loy, and X.~Tang, ``Accelerating the super-resolution
  convolutional neural network,'' in \emph{European conference on computer
  vision}.\hskip 1em plus 0.5em minus 0.4em\relax Springer, 2016, pp. 391--407.

\bibitem{shi2016real}
W.~Shi, J.~Caballero, F.~Husz{\'a}r, J.~Totz, A.~P. Aitken, R.~Bishop,
  D.~Rueckert, and Z.~Wang, ``Real-time single image and video super-resolution
  using an efficient sub-pixel convolutional neural network,'' in
  \emph{Proceedings of the IEEE conference on computer vision and pattern
  recognition}, 2016, pp. 1874--1883.

\bibitem{kim2016accurate}
J.~Kim, J.~Kwon~Lee, and K.~Mu~Lee, ``Accurate image super-resolution using
  very deep convolutional networks,'' in \emph{Proceedings of the IEEE
  conference on computer vision and pattern recognition}, 2016, pp. 1646--1654.

\bibitem{lim2017enhanced}
B.~Lim, S.~Son, H.~Kim, S.~Nah, and K.~Mu~Lee, ``Enhanced deep residual
  networks for single image super-resolution,'' in \emph{Proceedings of the
  IEEE conference on computer vision and pattern recognition workshops}, 2017,
  pp. 136--144.

\bibitem{tong2017image}
T.~Tong, G.~Li, X.~Liu, and Q.~Gao, ``Image super-resolution using dense skip
  connections,'' in \emph{Proceedings of the IEEE International Conference on
  Computer Vision}, 2017, pp. 4799--4807.

\bibitem{zhang2018image}
Y.~Zhang, K.~Li, K.~Li, L.~Wang, B.~Zhong, and Y.~Fu, ``Image super-resolution
  using very deep residual channel attention networks,'' in \emph{Proceedings
  of the European Conference on Computer Vision (ECCV)}, 2018, pp. 286--301.

\bibitem{dai2019second}
T.~Dai, J.~Cai, Y.~Zhang, S.-T. Xia, and L.~Zhang, ``Second-order attention
  network for single image super-resolution,'' in \emph{Proceedings of the IEEE
  Conference on Computer Vision and Pattern Recognition}, 2019, pp.
  11\,065--11\,074.

\bibitem{chen2019camera}
C.~Chen, Z.~Xiong, X.~Tian, Z.-J. Zha, and F.~Wu, ``Camera lens
  super-resolution,'' in \emph{Proceedings of the IEEE Conference on Computer
  Vision and Pattern Recognition}, 2019, pp. 1652--1660.

\bibitem{zhang2019zoom}
X.~Zhang, Q.~Chen, R.~Ng, and V.~Koltun, ``Zoom to learn, learn to zoom,'' in
  \emph{Proceedings of the IEEE Conference on Computer Vision and Pattern
  Recognition}, 2019, pp. 3762--3770.

\bibitem{cai2019toward}
J.~Cai, H.~Zeng, H.~Yong, Z.~Cao, and L.~Zhang, ``Toward real-world single
  image super-resolution: A new benchmark and a new model,'' \emph{arXiv
  preprint arXiv:1904.00523}, 2019.

\bibitem{zhang2018learning}
K.~Zhang, W.~Zuo, and L.~Zhang, ``Learning a single convolutional
  super-resolution network for multiple degradations,'' in \emph{Proceedings of
  the IEEE Conference on Computer Vision and Pattern Recognition}, 2018, pp.
  3262--3271.

\bibitem{wang2018esrgan}
X.~Wang, K.~Yu, S.~Wu, J.~Gu, Y.~Liu, C.~Dong, Y.~Qiao, and C.~Change~Loy,
  ``Esrgan: Enhanced super-resolution generative adversarial networks,'' in
  \emph{Proceedings of the European Conference on Computer Vision (ECCV)},
  2018, pp. 0--0.

\bibitem{bulat2018learn}
A.~Bulat, J.~Yang, and G.~Tzimiropoulos, ``To learn image super-resolution, use
  a gan to learn how to do image degradation first,'' in \emph{Proceedings of
  the European Conference on Computer Vision (ECCV)}, 2018, pp. 185--200.

\bibitem{zhou2019kernel}
R.~Zhou and S.~Susstrunk, ``Kernel modeling super-resolution on real
  low-resolution images,'' in \emph{Proceedings of the IEEE International
  Conference on Computer Vision}, 2019, pp. 2433--2443.

\bibitem{timofte2017ntire}
R.~Timofte, E.~Agustsson, L.~Van~Gool, M.-H. Yang, and L.~Zhang, ``Ntire 2017
  challenge on single image super-resolution: Methods and results,'' in
  \emph{Proceedings of the IEEE Conference on Computer Vision and Pattern
  Recognition Workshops}, 2017, pp. 114--125.

\bibitem{wengrowski2019light}
E.~Wengrowski and K.~J. Dana, ``Light field messaging with deep photographic
  steganography,'' in \emph{Proceedings of the IEEE conference on computer
  vision and pattern recognition}, 2019, pp. 1515--1524.

\bibitem{liu2019dual}
X.~Liu, M.~Suganuma, Z.~Sun, and T.~Okatani, ``Dual residual networks
  leveraging the potential of paired operations for image restoration,'' in
  \emph{Proceedings of the IEEE Conference on Computer Vision and Pattern
  Recognition}, 2019, pp. 7007--7016.

\bibitem{rosenfeld1976digital}
A.~Rosenfeld, \emph{Digital picture processing}.\hskip 1em plus 0.5em minus
  0.4em\relax Academic press, 1976.

\bibitem{bevilacqua2012low}
M.~Bevilacqua, A.~Roumy, C.~Guillemot, and M.~L. Alberi-Morel, ``Low-complexity
  single-image super-resolution based on nonnegative neighbor embedding,''
  2012.

\bibitem{zeyde2010single}
R.~Zeyde, M.~Elad, and M.~Protter, ``On single image scale-up using
  sparse-representations,'' in \emph{International conference on curves and
  surfaces}.\hskip 1em plus 0.5em minus 0.4em\relax Springer, 2010, pp.
  711--730.

\bibitem{arbelaez2010contour}
P.~Arbelaez, M.~Maire, C.~Fowlkes, and J.~Malik, ``Contour detection and
  hierarchical image segmentation,'' \emph{IEEE transactions on pattern
  analysis and machine intelligence}, vol.~33, no.~5, pp. 898--916, 2010.

\bibitem{aly2005image}
H.~A. Aly and E.~Dubois, ``Image up-sampling using total-variation
  regularization with a new observation model,'' \emph{IEEE Transactions on
  Image Processing}, vol.~14, no.~10, pp. 1647--1659, 2005.

\bibitem{xiong2010robust}
Z.~Xiong, X.~Sun, and F.~Wu, ``Robust web image/video super-resolution,''
  \emph{IEEE transactions on image processing}, vol.~19, no.~8, pp. 2017--2028,
  2010.

\bibitem{he2013beta}
L.~He, H.~Qi, and R.~Zaretzki, ``Beta process joint dictionary learning for
  coupled feature spaces with application to single image super-resolution,''
  in \emph{Proceedings of the IEEE conference on computer vision and pattern
  recognition}, 2013, pp. 345--352.

\bibitem{freeman2002example}
W.~T. Freeman, T.~R. Jones, and E.~C. Pasztor, ``Example-based
  super-resolution,'' \emph{IEEE Computer graphics and Applications}, no.~2,
  pp. 56--65, 2002.

\bibitem{kim2016deeply}
J.~Kim, J.~Kwon~Lee, and K.~Mu~Lee, ``Deeply-recursive convolutional network
  for image super-resolution,'' in \emph{Proceedings of the IEEE conference on
  computer vision and pattern recognition}, 2016, pp. 1637--1645.

\bibitem{ledig2017photo}
C.~Ledig, L.~Theis, F.~Husz{\'a}r, J.~Caballero, A.~Cunningham, A.~Acosta,
  A.~Aitken, A.~Tejani, J.~Totz, Z.~Wang \emph{et~al.}, ``Photo-realistic
  single image super-resolution using a generative adversarial network,'' in
  \emph{Proceedings of the IEEE conference on computer vision and pattern
  recognition}, 2017, pp. 4681--4690.

\bibitem{johnson2016perceptual}
J.~Johnson, A.~Alahi, and L.~Feifei, ``Perceptual losses for real-time style
  transfer and super-resolution,'' in \emph{Proceedings of the European
  Conference on Computer Vision (ECCV)}, 2016, pp. 694--711.

\bibitem{efrat2013accurate}
N.~Efrat, D.~Glasner, A.~Apartsin, B.~Nadler, and A.~Levin, ``Accurate blur
  models vs. image priors in single image super-resolution,'' in
  \emph{Proceedings of the IEEE International Conference on Computer Vision},
  2013, pp. 2832--2839.

\bibitem{simonyan2014very}
K.~Simonyan and A.~Zisserman, ``Very deep convolutional networks for
  large-scale image recognition,'' in \emph{ICLR}, 2015.

\bibitem{szegedy2015going}
C.~Szegedy, W.~Liu, Y.~Jia, P.~Sermanet, S.~Reed, D.~Anguelov, D.~Erhan,
  V.~Vanhoucke, and A.~Rabinovich, ``Going deeper with convolutions,'' in
  \emph{Proceedings of the IEEE conference on computer vision and pattern
  recognition}, 2015, pp. 1--9.

\bibitem{he2016deep}
K.~He, X.~Zhang, S.~Ren, and J.~Sun, ``Deep residual learning for image
  recognition,'' in \emph{Proceedings of the IEEE conference on computer vision
  and pattern recognition}, 2016, pp. 770--778.

\bibitem{krizhevsky2017imagenet}
A.~Krizhevsky, I.~Sutskever, and G.~E. Hinton, ``Imagenet classification with
  deep convolutional neural networks,'' \emph{Communications of The ACM},
  vol.~60, no.~6, pp. 84--90, 2017.

\bibitem{huang2017densely}
G.~Huang, Z.~Liu, L.~V. Der~Maaten, and K.~Q. Weinberger, ``Densely connected
  convolutional networks,'' in \emph{Proceedings of the IEEE conference on
  computer vision and pattern recognition}, 2017, pp. 2261--2269.

\bibitem{he2017mask}
K.~He, G.~Gkioxari, P.~Dollar, and R.~Girshick, ``Mask r-cnn,'' in
  \emph{Proceedings of the IEEE conference on computer vision and pattern
  recognition}, 2017, pp. 2980--2988.

\bibitem{haris2018deep}
M.~Haris, G.~Shakhnarovich, and N.~Ukita, ``Deep back-projection networks for
  super-resolution,'' in \emph{Proceedings of the IEEE conference on computer
  vision and pattern recognition}, 2018, pp. 1664--1673.

\bibitem{hu2019squeeze-and-excitation}
J.~Hu, L.~Shen, S.~Albanie, G.~Sun, and E.~Wu, ``Squeeze-and-excitation
  networks,'' \emph{IEEE Transactions on Pattern Analysis and Machine
  Intelligence}, pp. 1--1, 2019.

\bibitem{8100166}
F.~{Wang}, M.~{Jiang}, C.~{Qian}, S.~{Yang}, C.~{Li}, H.~{Zhang}, X.~{Wang},
  and X.~{Tang}, in \emph{Proceedings of the IEEE conference on computer vision
  and pattern recognition}, 2017, pp. 6450--6458.

\bibitem{woo2018cbam}
S.~Woo, J.~Park, J.~Lee, and I.~S. Kweon, ``Cbam: Convolutional block attention
  module,'' pp. 3--19, 2018.

\bibitem{vaswani2017attention}
A.~Vaswani, N.~Shazeer, N.~Parmar, J.~Uszkoreit, L.~Jones, A.~N. Gomez,
  L.~Kaiser, and I.~Polosukhin, ``Attention is all you need,'' pp. 5998--6008,
  2017.

\bibitem{lowe2004distinctive}
D.~G. Lowe, ``Distinctive image features from scale-invariant keypoints,''
  \emph{International journal of computer vision}, vol.~60, no.~2, pp. 91--110,
  2004.

\bibitem{fischler1981random}
M.~A. Fischler and R.~C. Bolles, ``Random sample consensus: a paradigm for
  model fitting with applications to image analysis and automated
  cartography,'' \emph{Communications of the ACM}, vol.~24, no.~6, pp.
  381--395, 1981.

\bibitem{zhang2019deep}
K.~Zhang, W.~Zuo, and L.~Zhang, ``Deep plug-and-play super-resolution for
  arbitrary blur kernels,'' in \emph{Proceedings of the IEEE conference on
  computer vision and pattern recognition}, 2019, pp. 1671--1681.

\bibitem{gu2019blind}
J.~Gu, H.~Lu, W.~Zuo, and C.~Dong, ``Blind super-resolution with iterative
  kernel correction,'' in \emph{Proceedings of the IEEE conference on computer
  vision and pattern recognition}, 2019, pp. 1604--1613.

\bibitem{jolicoeurmartineau2019the}
A.~{Jolicoeur-Martineau}, ``The relativistic discriminator: a key element
  missing from standard gan,'' in \emph{ICLR 2019 : 7th International
  Conference on Learning Representations}, 2019.

\bibitem{goodfellow2014generative}
I.~Goodfellow, J.~Pougetabadie, M.~Mirza, B.~Xu, D.~Wardefarley, S.~Ozair,
  A.~Courville, and Y.~Bengio, ``Generative adversarial nets,'' pp. 2672--2680,
  2014.

\bibitem{zhang2018residual}
Y.~Zhang, Y.~Tian, Y.~Kong, B.~Zhong, and Y.~Fu, ``Residual dense network for
  image super-resolution,'' in \emph{Proceedings of the IEEE Conference on
  Computer Vision and Pattern Recognition}, 2018, pp. 2472--2481.

\bibitem{lai2017deep}
W.-S. {Lai}, J.-B. {Huang}, N.~{Ahuja}, and M.-H. {Yang}, ``Deep laplacian
  pyramid networks for fast and accurate super-resolution,'' in \emph{2017 IEEE
  Conference on Computer Vision and Pattern Recognition (CVPR)}, 2017, pp.
  5835--5843.

\bibitem{muller2019does}
R.~M{\"u}ller, S.~Kornblith, and G.~E. Hinton, ``When does label smoothing
  help?'' in \emph{Advances in Neural Information Processing Systems}, 2019,
  pp. 4696--4705.

\bibitem{wang2004image}
Z.~Wang, A.~C. Bovik, H.~R. Sheikh, E.~P. Simoncelli \emph{et~al.}, ``Image
  quality assessment: from error visibility to structural similarity,''
  \emph{IEEE transactions on image processing}, vol.~13, no.~4, pp. 600--612,
  2004.

\bibitem{kingma2014adam}
D.~P. Kingma and J.~Ba, ``Adam: A method for stochastic optimization,''
  \emph{In ICLR}, 2014.

\bibitem{gonzalez2004digital}
R.~C. Gonzalez, R.~E. Woods, and S.~L. Eddins, \emph{Digital image processing
  using MATLAB}.\hskip 1em plus 0.5em minus 0.4em\relax Pearson Education
  India, 2004.

\end{thebibliography}

%

\end{document}